\documentclass[amsmath,10pt,twocolumn,superscriptaddress,footinbib,aip,reprint]{revtex4-2}
\usepackage{amsmath,amsfonts,bm,dsfont}
\usepackage{graphicx}
\usepackage{subfigure}
\usepackage{float}
\usepackage{comment}
\usepackage{chemformula}
\usepackage[normalem]{ulem}
\usepackage{physics}
\usepackage{extpfeil}
\usepackage{uri}

\newcommand{\appropto}{\mathrel{\vcenter{
  \offinterlineskip\halign{\hfil$##$\cr
    \propto\cr\noalign{\kern2pt}\sim\cr\noalign{\kern-2pt}}}}}

\usepackage{color}
\usepackage[colorlinks=true,allcolors=blue]{hyperref}
\usepackage{tikz}
\usepackage{upgreek}
\usepackage{amssymb}

\definecolor{rxncolor}{rgb}{0.141176, 0.529412, 0.129412}
\definecolor{mpocolor}{rgb}{0.777778, 0.555556, 0.777778}
\definecolor{Acolor}{rgb}{0.531316, 0.685913, 0.947556}
\definecolor{A2color}{rgb}{0.353605, 0.462894, 0.92237}
\definecolor{OA2color}{rgb}{0.213307, 0.226237, 0.848147}
\definecolor{Ocolor}{rgb}{0.0, 0.0, 0.0}
\definecolor{OB2color}{rgb}{0.617469, 0.171738, 0.222554}
\definecolor{B2color}{rgb}{0.759441, 0.374849, 0.348673}
\definecolor{Bcolor}{rgb}{0.855666, 0.590976, 0.489811}

\DeclareUnicodeCharacter{0301}{\'{e}}

\begin{document}
\title{Chemical master equation parameter exploration using DMRG}
\author{John P.~Zima}
\author{Schuyler B.~Nicholson}
\author{Todd R.~Gingrich}
\email{todd.gingrich@northwestern.edu.}
\affiliation{Department of Chemistry, Northwestern University, 2145 Sheridan Road, Evanston, Illinois 60208, USA}

\begin{abstract}
  Well-mixed chemical reaction networks (CRNs) contain many distinct chemical species with copy numbers that fluctuate in correlated ways.
  While those correlations are typically monitored via Monte Carlo sampling of stochastic trajectories, there is interest in systematically approximating the joint distribution over the exponentially large number of possible microstates using tensor networks or tensor trains.
  We exploit the tensor network strategy to determine when the steady state of a seven-species gene toggle switch CRN model supports bistability as a function of two decomposition rates, both parameters of the kinetic model.
  We highlight how the tensor network solution captures the effects of stochastic fluctuations, going beyond mean field and indeed deviating meaningfully from a mean-field analysis.
  The work furthermore develops and demonstrates several technical advances that will allow steady-states of broad classes of CRNs to be computed in a manner conducive to parameter exploration.
  We show that the steady-state distributions can be computed via the ordinary density matrix renormalization group (DMRG) algorithm despite having a non-Hermitian rate operator with a small spectral gap, we illustrate how that steady-state distribution can be efficiently projected to an order parameter that identifies bimodality, and we employ excited-state DMRG to calculate a relaxation timescale for the bistability.
\end{abstract}
\maketitle
\section{Introduction}
\label{sec:intro}

Markovian master equations appear in diverse disciplines~\cite{goutsias2013markovian}.
Though the models are ubiquitous, closed-form analytical solutions are rare and limited to very simplified models.
Complex systems therefore often require Monte Carlo sampling of representative stochastic trajectories.
Unfortunately, na\"{i}ve sampling methods like the Gillespie algorithm~\cite{gillespie1977exact} can suffer from slow convergence, particularly when used to compute properties of systems with a separation of timescales between frequent and rare events.
In that case, for a rare event to be observed, orders of magnitude more frequent events must first be sampled.
Generating an outsized sampling of frequent events can be prohibitively costly, especially because the typical time steps taken by the Gillespie algorithm scale inversely with system size.

The computational expense is dramatically amplified when one wants to solve inverse problems.
Then one hopes to explore the forward problem\textemdash given a model and its rate parameters, what stochastic dynamics emerges\textemdash for a large number of parameters~\cite{goutsias2013markovian, minas2019parameter}, demanding ways to efficiently extract the sensitivity to the parameters~\cite{rijal2024differentiable}.
Ideally, one should leverage the computational effort invested for one parameter set to more cheaply study the subsequent parameter sets, but brute-force Gillespie sampling does not easily lend itself to cost savings.
Reweighting samples from one set of system parameters to provide insight about a different but similar set of parameters costs essentially the same effort as resampling from scratch.

When the forward problem features rapid but rare events, there are enhanced sampling techniques that focus on slow-timescale transitions between different metastable regions of configuration space.
Transition path sampling (TPS)~\cite{bolhuis2002transition, crooks2001efficient, gingrich2015preserving} and forward flux sampling (FFS)~\cite{allen2005sampling, allen2006simulating, allen2009forward} can both generate representative trajectories of those transitions, but both classes of methods bring other complications.
TPS requires an initial transition path and a sometimes fickle set of Monte Carlo moves to hop from one reactive trajectory to another; FFS requires identification of a ``good'' low-dimensional reaction coordinate.

Given the challenges, it is worthwhile to reconsider whether sampling approaches are truly necessary.
Could one bypass the realizations and instead compute the full steady-state distribution?
Initially, this idea seems impractical.
The number of microstates scales exponentially with system size, and calculations involving the steady-state distribution require that one compute the probability for each of those exponentially many microstates.
Suppose, however, that one sought instead a controllable approximation to that distribution.
Due to the modern success of machine learning, one might look to neural networks as the natural approximation tool.
Perhaps one could use stochastic gradient descent to fit a billion- or trillion-parameter nonlinear function to a large number of sampled trajectories, for example.

As we detail in this work, we can achieve a parameterized, controllable approximation to the distribution while avoiding both trajectory sampling and stochastic gradient descent.
However, the key is not the use of a neural network.
Rather, we fit a high-dimensional \emph{multi-linear} function using a \emph{non-gradient} optimization procedure that draws directly from the rules of the stochastic dynamics (the rates in the master equation), not from sampled data.
The multilinear functions to be fitted are graphically represented as tensor networks~\cite{orus2014practical, orus2019tensor}, most often a one-dimensional chain of contracted tensors known in physics communities as a matrix product state (MPS) and in the applied math community as a tensor train (TT) \cite{oseledets2011tensor}.
Parameterization of the distribution boils down to fitting the elements of all tensors in the chain, an optimization problem that can be performed by sweeping through the tensors and optimizing a single tensor at a time.
The core tensor network ideas and optimization strategies are very well developed and broadly known in the quantum many-body \cite{schollwock2011density, orus2014practical, orus2019tensor, white1992density, white1993density}, quantum chemistry \cite{chan2011density, wouters2014density, sharma2012spin, white1999ab, chan2002highly}, and numerical linear algebra \cite{kolda2009tensor, oseledets2011tensor, grasedyck2013literature, dolgov2014alternating, dolgov2015simultaneous} communities.
They have begun to be used for classical master equations~\cite{garrahan2016classical, helms2019dynamical, strand2022using, strand2022computing, garrahan2024topological, causer2021optimal, causer2022finite, causer2020dynamics, causer2023optimal, causer2022slow, merbis2024effective, dolgov2024tensor}, including applications for chemical master equations (CMEs)~\cite{hegland2010numerical, kazeev2014direct, dolgov2015simultaneous, liao2015tensor, gelss2016solving, vo2017adaptive, ion2021tensor, nicholson2023quantifying}.
An advantage of these techniques over sampling-based approaches lies in the fact that once the distribution has been approximated for one parameter set, that approximated distribution can be used as a very effective seed when studying parameters with a similar steady-state distribution.

Gelß et al.\ leveraged this advantage when studying a heterogeneous catalysis model of CO oxidation on a RuO$_2$(110) surface~\cite{gelss2016solving, holtz2012alternating}.
As the master equation stiffened, rare events became rarer and the computational cost of kinetic Monte Carlo (Gillespie) scaled worse than the cost of converging an MPS approximation using the so-called Alternating Linear Scheme (ALS)~\cite{holtz2012alternating}.
Liao et al.\ have made significant strides in applying this class of methods to biochemical kinetics, tackling multiple chemical reaction networks, including a cell cycle model~\cite{liao2015tensor}. Their approach uses tensor networks to solve the chemical Fokker-Planck equation (CFPE) that arises from truncating the Kramers-Moyal expansion of the CME at second order. They discretized the problem on a grid and approximated the steady state by converging an MPS with the Alternating Minimal Energy Method (AMEn)~\cite{dolgov2014alternating}.
The procedure is flexible in that it can be performed with different discretized grids, but passing from the CME to the CFPE is problematic in the low-copy-number regime if the particular CME exhibits ``discreteness-induced multimodality''~\cite{duncan2015noise, erban2020stochastic}.
That low-copy-number regime is essential for some kinetic models of biological processes.
Take, for instance, models of a gene toggle switch (GTS)~\cite{warren2004enhancement} with fluctuating amounts of proteins A and B, each suppressing the activation of the gene that produces the other protein.
 With many copies of DNA, proteins A and B coexist since, at any given time, some DNA molecules would be in an A-producing state and others in a B-producing state.
Inside a single cell, however, one cannot ignore the discreteness of the DNA, which could cause the cell to flip between A- and B-rich states.

To handle both small- and large-copy-number regimes, we build the tensor network approach directly from the CME using the Doi-Peliti formalism~\cite{doi1976stochastic,peliti1985path,nicholson2023quantifying}.
This formalism, a classical analog of second quantization methods, allows a set of elementary reactions to be written in terms of raising and lowering operators that act on each species.
By writing each reaction of the GTS model in this second-quantized form, the rate operator that generates the CME dynamics can be numerically constructed as a matrix product operator (MPO) using standard software packages \cite{10.21468/SciPostPhysCodeb.4, 10.21468/SciPostPhysCodeb.4-r0.3}.
The steady-state distribution over microstates is the top eigenvector of that MPO.
 In other words, solving for the steady-state distribution is analogous to using the Density Matrix Renormalization Group (DMRG) algorithm to find the ground state of a Hamiltonian.
 Unlike in the usual quantum mechanical context, the rate operator MPO is typically not Hermitian, but we appeal to the Time Dependent Variational Principle (TDVP) to rationalize why the ordinary DMRG algorithm can nevertheless compute the steady state.
 That steady-state distribution can be easily marginalized to obtain a distribution for an order parameter that detects bimodality, allowing for the efficient construction of a kinetic phase diagram.

 The paper is structured as follows.
 Section~\ref{sec:CRN&GTS} introduces chemical reaction networks and the chemical master equation, specializing in the case of the exclusive gene toggle switch model of Ref.~\cite{warren2004enhancement}.
 Section~\ref{sec:TNs} reviews the tensor network tools, both the MPS used for the distribution and the MPO used for the rate operator.
 Section~\ref{sec:DMRG} discusses the DMRG algorithm, highlighting why standard DMRG algorithms can target the steady-state distribution for CMEs even though the time propagator is not Hermitian.
 Section~\ref{sec:bistability} shows how to use the tensor network tools to assess the bimodality and bistability of the CME.
 Section~\ref{sec:exploration} presents our main numerical results, demonstrating how reseeded DRMG calculations can help explore the parameter space.
 We close with a discussion of the outlook and challenges.

\section{Chemical Reaction Networks and the Gene Toggle Switch Model}
\label{sec:CRN&GTS}

\begin{figure*}
\centering
\includegraphics[width=0.9\textwidth]{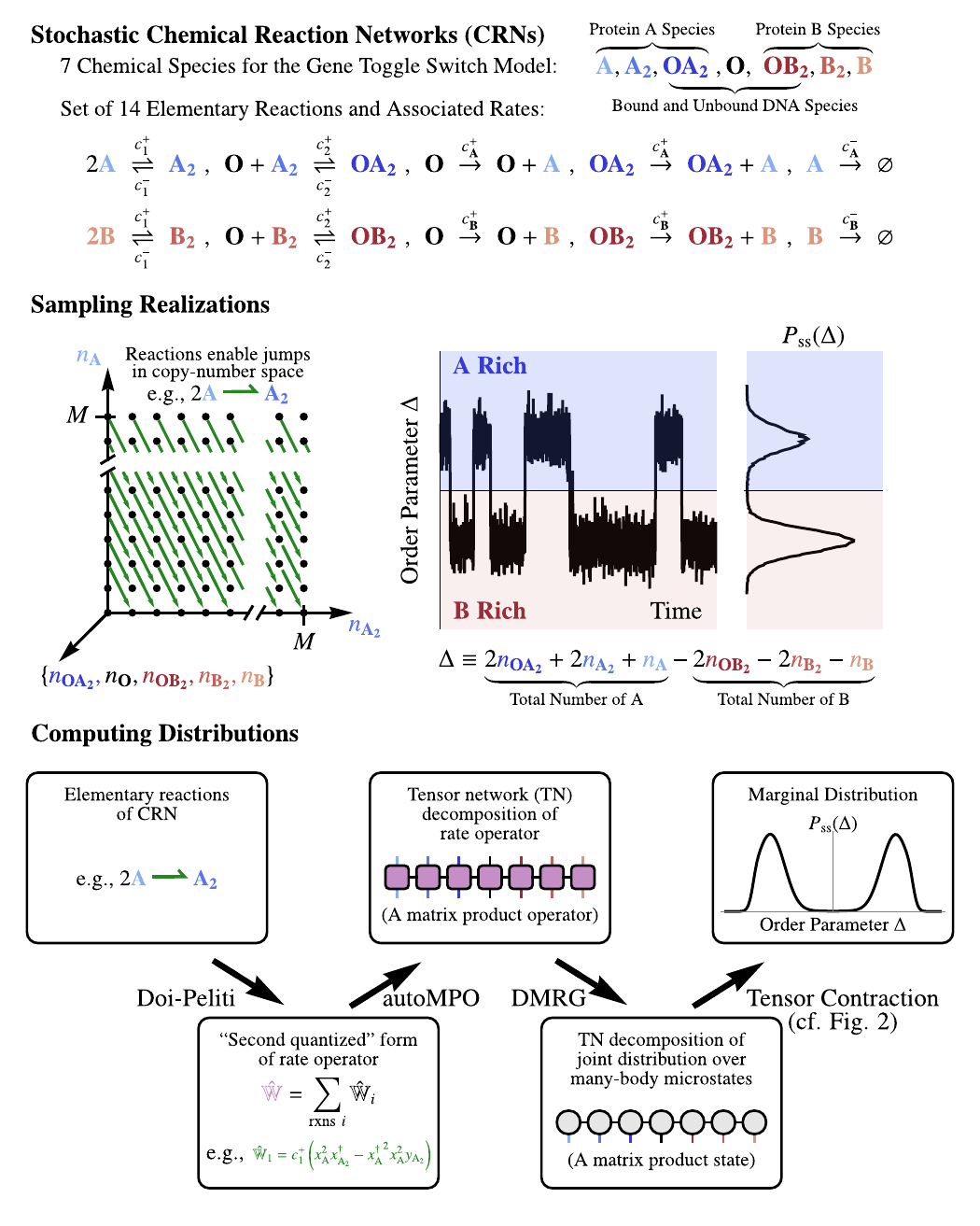}
\caption{\label{fig:overall_scheme}
  Two approaches to stochastic chemical kinetics, illustrated with the gene toggle switch (GTS) model.
  {\bf (Top)} The 14 elementary reactions of the model couple together fluctuations in the copy number of the seven chemical species.
  The model is especially notable for generating a bistable switch.
  {\bf (Middle)}
  Traditionally, the time-evolution for those copy-number fluctuations is sampled by a kinetic Monte Carlo (Gillespie) trajectory that jumps between a set of microstates too numerous to practically enumerate.
  With sufficiently long trajectories, one can build converged histograms for low-dimensional order parameters like \(\Delta\) to sample \(P_{\rm ss}(\Delta)\).
  {\bf (Bottom)}
  This manuscript describes an approach to numerically compute those same distributions without sampling.
  Rather, the elementary reactions are converted into a raising and lowering operator form (Doi-Peliti), allowing the chemical master equation to be expressed in terms of a tensor network operator.
  The steady-state distribution can be controllably approximated as a matrix product state.}
\end{figure*}

In this paper, we focus on well-mixed chemical reaction networks (CRNs), which are comprised of a set of \(N\) species and \(K\) elementary reactions.
This paper is organized around a specific example CRN, the GTS model.
In that example, the 7 species and 14 reactions are depicted explicitly in Fig.~\ref{fig:overall_scheme}.
CRNs like the GTS model become rich and interesting because the outputs of some elementary reactions are inputs for others, coupling together the elementary steps into a complex kinetic network.
The dynamics of the network is monitored by tracking the microstate of the CRN, specified by an \(N\)-element vector \({\bf n}\) whose elements give the copy number of each species.
For GTS, ${\bf n} = (\color{Acolor} n_{\rm A} \color{Ocolor}, \color{A2color} n_{\rm A_2} \color{Ocolor}, \color{OA2color} n_{\rm OA_2} \color{Ocolor}, \color{Ocolor} n_{\rm O} \color{Ocolor}, \color{OB2color} n_{\rm OB_2} \color{Ocolor}, \color{B2color} n_{\rm B_2} \color{Ocolor}, \color{Bcolor} n_{\rm B} \color{Ocolor})$.
The action of elementary reaction \(k\) is encoded in an \(N\)-element vector \(\boldsymbol{\delta} {\bf n}_k\) that reports the change in copy number for each species when reaction \(k\) fires.
Let reaction 1 be \(2 \color{Acolor} \text{A} \color{Ocolor} \xrightharpoonup{} \color{A2color} \text{A}_2\) as in Fig.~\ref{fig:overall_scheme}.
Then \(\boldsymbol{\delta} {\bf n}_1 = (-2, 1, 0, 0, 0, 0, 0)\).
The propensity for a particular reaction to fire is a product of two terms, a combinatorial term counting the number of unique sets of reaction \(k\)'s reactants and a rate for any one of those sets to react.
This propensity \(w_k({\bf n})\) depends on the microstate, since the combinatorial factor depends on \({\bf n}\).
The probability of occupying that microstate at time \(t\) evolves according to the chemical master equation (CME):
\begin{equation}
\frac{dp(\mathbf{n}, t)}{dt} = \sum_{k=1}^K \left[ w_k(\mathbf{n} - \delta \mathbf{n}_k) p(\mathbf{n} - \delta \mathbf{n}_k, t) - w_k(\mathbf{n}) p(\mathbf{n}, t) \right],
\label{eq:CME}
\end{equation}
with the first term of the summand giving the rate of gaining probability in microstate \({\bf n}\) from reaction \(k\) and the second term giving the rate of losing probability from \({\bf n}\) due to that same elementary reaction.

By choosing an ordering of microstates, Eq.~\eqref{eq:CME} can be cast in matrix form.
We combine the probabilities \(p({\bf n}, t)\) into a vector \(\left|p(t)\right>\) with component \(i\) corresponding to the probability of the \(i^{\rm th}\) microstate at time \(t\), and we define the rate matrix \(\mathbb{W}\) to have off-diagonal elements \(\mathbb{W}_{ij}\) that sum \(w_k\) over all reactions \(k\) which transform microstate \(j\) into \(i\).
The off-diagonal elements of \(\mathbb{W}\) capture the ``gain'' terms of Eq.~\eqref{eq:CME}.
The negative ``loss'' terms are accounted for by setting diagonal terms such that the columns sum to zero: \(\mathbb{W}_{ii} = -\sum_{j \neq i} \mathbb{W}_{ji}\), a condition that ensures conservation of probability.
Eq.~\eqref{eq:CME} then becomes simply
\begin{equation}
\frac{d\left|p(t)\right>}{dt} = \mathbb{W} \left|p(t)\right>.
\end{equation}
This master equation is formally solved by \(\left|p(t)\right> = e^{\mathbb{W} t} \left|p(0)\right>\).
Under the relatively weak condition of irreducibility (all microstates can be reached by mixing together a sequence of elementary reactions), the long-time steady-state probability distribution is given by the top right eigenvector of \({\mathbb{W}}\).
The simplicity of the formal solutions to the master equation hides the complexity brought by the particular structure of \(\mathbb{W}\) and the need to compute spectral properties of that very large matrix.

We emphasize this challenge with our focus on the GTS of Warren and ten Wolde~\cite{warren2004enhancement}, chosen as the illustrative CRN for three reasons.
Firstly, the model is well understood, having been studied using mean field, Gillespie, and even advanced sampling procedures~\cite{allen2005sampling}.
Secondly, while the model's state space is too large to treat directly with dense linear algebra, it sits at the boundary of what is possible using sparse matrix methods~\cite{strand2024from}, offering a useful comparison when assessing errors (see Appendix~\ref{app:fourth}).
Thirdly, the model supports a regime where the discreteness of the species' copy numbers is materially important, making GTS a challenging test of methodology.

The GTS model comprises seven species and fourteen reactions, as shown in Fig.~\ref{fig:overall_scheme}.
Two of those species, $\mathrm{A}$ and $\mathrm{B}$, represent proteins which can be synthesized by DNA and which decompose through different elementary reactions with rates \(c_{\rm A}^-\) and \(c_{\rm B}^-\).
A defining characteristic of the GTS model is that both proteins suppress the production of the other.
That production requires DNA, which exists in one of three forms: unbound (species O), bound to a dimer of protein A (species OA\(_2\)), or bound to a dimer of protein B (species OB\(_2\)).
All three forms can synthesize protein.
Although O can synthesize both A and B, OA\(_2\) can only synthesize A, and OB\(_2\) can only synthesize B.
The final two species are the protein dimers A\(_2\) and B\(_2\), which reversibly bind to the DNA and which reversibly form from the respective monomers.
These rules and the associated rate constants of Fig.~\ref{fig:overall_scheme} are sufficient to prescribe the form of \(\mathbb{W}\), but that does not mean that it is trivial to practically construct the matrix or compute the steady state.

The core problem is the curse of dimensionality.
A microstate of the model is given by the copy numbers $(\color{Acolor} n_{\rm A} \color{Ocolor}, \color{A2color} n_{\rm A_2} \color{Ocolor}, \color{OA2color} n_{\rm OA_2} \color{Ocolor}, \color{Ocolor} n_{\rm O} \color{Ocolor}, \color{OB2color} n_{\rm OB_2} \color{Ocolor}, \color{B2color} n_{\rm B_2} \color{Ocolor}, \color{Bcolor} n_{\rm B} \color{Ocolor})$.
Each copy number must be a nonnegative counting number, so microstates are lattice points in a seven-dimensional space.
Conservation laws in CRNs can slightly reduce the size of that space.
For example, none of the 14 elementary reactions creates or destroys DNA; they only interconvert between its three forms (O, OA\(_2\), and OB\(_2\)).
As a consequence, the dynamics is only irreducible within a manifold of states with a conserved value of \(n_{\rm DNA} \equiv \color{OA2color} n_{\rm OA_2} \color{Ocolor} + n_{\rm O} +  \color{OB2color} n_{\rm OB_2}\).
In this paper, we study the \(n_{\rm DNA} = 1\), single-DNA regime, which means that microstates span a five-dimensional space.
That single DNA exists in one of three states, but the copy numbers of A, A\(_2\), B, and B\(_2\) can all vary from 0 to \(\infty\).
Gillespie sampling executes a random walk on that infinite set of copy numbers, randomly drawing a next reaction and following that reaction to get to the next lattice point (see Fig.~\ref{fig:overall_scheme}).
Though the random walk occurs in the high-dimensional space of all copy numbers, the trajectory can be projected down onto a lower dimensional order parameter.
For GTS, it is natural to look at the evolution of \(\Delta \equiv 2 \color{OA2color} n_{\mathrm{OA_2}} \color{Ocolor} + 2 \color{A2color} n_{\mathrm{A_2}} \color{Ocolor} + \color{Acolor} n_{\mathrm{A}} \color{Ocolor} - 2 \color{OB2color} n_{\mathrm{OB_2}} \color{Ocolor} - 2 \color{B2color} n_{\mathrm{B_2}} \color{Ocolor} - \color{Bcolor} n_{\mathrm{B}}\), which counts the excess of protein A relative to protein B.
Fig.~\ref{fig:overall_scheme} shows a representative Gillespie trajectory that illustrates the toggle-switch nature of the GTS model; the system switches between two metastable macrostates, an A-rich state characterized by positive \(\Delta\) and a B-rich state with negative \(\Delta\).
Rare, stochastic fluctuations trigger the flips between macrostates.
With enough sampling, the steady-state distribution for the order parameter, \(P_{\rm ss}(\Delta)\), can be constructed as a histogram, but many of the rare flips would need to be observed to converge.

An alternative way to study the GTS is to numerically compute the steady-state distribution from the CME without sampling.
Doing so clearly requires a truncation to allow only a finite number of microstates.
By disallowing copy numbers for A, A\(_2\), B, and B\(_2\) from growing larger than \(M\), we limit the number of microstates (and thus the dimension of the rate matrix \(\mathbb{W}\)) to \(3 (M+1)^4\).
For the elementary-reaction rate constants studied in this work, the protein degradation rates \(c_{\rm A}^-\) and \(c_{\rm B}^-\) were sufficiently large that \(M = 20\) was a reasonable truncation.
Incorporating this truncation and the \(n_{\rm DNA} = 1\) conservation law, numerically solving for \(\left|\pi\right> \equiv \lim_{t \to \infty} \left|p(t)\right>\) amounts to finding the largest right eigenvector of a sparse \(\mathbb{W}\) with dimension 583,433.
That problem is accessible to sparse linear algebra methods like the generalized minimal residual method (GMRES), but it is easy to anticipate how increases in \(M\) or the number of species could quickly overwhelm such numerical procedures.
We now move on to introduce the tensor network tools that offer a powerful and controlled approximation of \(\left|\pi\right>\), which generalizes to cases where it is not feasible to enumerate microstates \cite{nicholson2023quantifying}.

\section{Tensor Networks}
\label{sec:TNs}

If one is to break through the curse of dimensionality and calculate \(\left|\pi\right>\) for a CME with an astronomically large number of microstates, it is imperative to shift from thinking of the steady state as being given by the vector \(\left|\pi\right>\) and instead thinking of the steady state as a function of microstates \(\pi(\mathbf{n})\).
When \(\left|\pi\right>\) is a vector, it is necessary to independently list the probability of every microstate to fully define \(\left|\pi\right>\).
Solving for the steady-state distribution means solving for each of those elements of the vector.
By instead viewing the steady-state distribution as a function of \(\mathbf{n}\), we can ask if we could controllably and accurately approximate the true function \(\pi(\mathbf{n})\) by a simpler function.
Approximating the steady-state distribution then requires that one fit that simpler function, potentially with a very large number of fitting parameters.
To be controllable, the approximation must offer a simple way to increase accuracy at the expense of more computation.

A very appealing controllable approximation is to choose multi-linear functions.
This choice is not a statement that the CRN can only involve ``linear reactions''.
Rather, it is a reflection of the fact that multi-linear functions can become very expressive if they are high dimensional and involve sufficiently many fitting parameters.
By working with multi-linear functions, we have the enormous advantage that we can deploy methods of numerical linear algebra, ultimately offering a route to efficiently fit the function.
These multi-linear functions are graphically represented as networks of tensors, and there is a very large literature devoted to the study and application of those tensor networks (TNs).
We refer interested readers to some fantastic pedagogical introductions~\cite{orus2014practical, orus2019tensor}.
In an attempt to make this paper self-contained and accessible to a broad audience, we provide a spartan overview of the graphical notation used in the field.
For our purposes, tensors are simply multi-dimensional arrays, with the order of the tensor being the number of dimensions in the array\textemdash a scalar is order 0, a vector order 1, a matrix order 2, etc.
Tensors are graphically represented by geometric shapes (here circles, rectangles, and rounded rectangles are all tensors), and the number of legs protruding from the shape tells the order of the tensor.
Each of those legs carries information about the state of one of the dimensions of the tensor.
When the legs are dangling, as in
\centerline{\includegraphics[height=50pt,trim=0.6cm 0.25cm 0.6cm 0.25cm,clip]{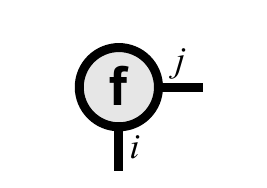},}
the legs should be thought of as inputs to the function \(f\).
If you specify the value of the index \(i\) and the value of the index \(j\), \(f_{ij}\) returns a number.
Networks of tensors are constructed when a leg of one tensor is connected to that of another to graphically denote tensor contraction.
For example,
\centerline{\includegraphics[height=50pt,trim=0.6cm 0cm 0.6cm 0cm,clip]{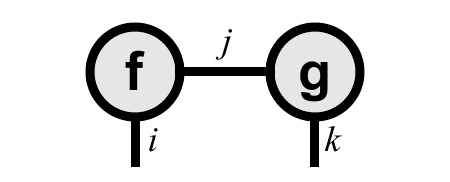},}
is a very simple TN.
The connection of the legs indicates that the values of the index must match, and the notation carries an implied summation over all possible values of that matching index, e.g. \(\sum_{j} f_{ij} g_{jk}\).
In words, this simple TN illustrates the multiplication of a matrix \(f\) with another \(g\) to yield a third matrix (two dangling legs remain).
Suppose one has a steady-state distribution \(\pi(i, k)\) which depended on the state of only two species with copy numbers given by \(i\) and \(k\).
The function \(\pi\) could be specified by determining \(\pi_{ik}\) for each of the microstates, but perhaps \(\pi\) could be approximated via the preceding TN in terms of \(f\) and \(g\).
The expense and accuracy of that approximation depends on the dimensionality of the leg connecting \(f\) to \(g\), the so-called bond dimension.
In other words, the sum over index \(j\) ranges from 1 up to the bond dimension.
When the bond dimension is large, \(f\) and \(g\) are larger matrices fit by more parameters, and the approximating function is more expressive.
In fact, if the bond dimension were sufficiently large, one could build \(\pi\) exactly as a product of a matrix \(f\) and \(g\) with no approximation.
Having demonstrated the graphical idea on the simplest of TNs, we are finally in a position to approximate \(\left|\pi\right>\) for the GTS model.

\subsection{Probability Distributions in a Matrix Product State Form}
\label{sec:MPS}

A distribution over microstates takes the form
\begin{equation}
  \left|p\right> = \sum_{\mathbf{n}} p_{\color{Acolor} n_{\rm A} \color{Ocolor}, \color{A2color} n_{\rm A_2} \color{Ocolor}, \color{OA2color} n_{\rm OA_2} \color{Ocolor}, \color{Ocolor} n_{\rm O} \color{Ocolor}, \color{OB2color} n_{\rm OB_2} \color{Ocolor}, \color{B2color} n_{\rm B_2} \color{Ocolor}, \color{Bcolor} n_{\rm B} \color{Ocolor}} \left|\mathbf{n}\right>,
\end{equation}
where \(\left|\mathbf{n}\right>\) is a basis vector corresponding to state \(\mathbf{n}\).
The sum is over all microstates, and \(p\) can be seen as an order seven tensor.
Given the occupation number of each species, \((\color{Acolor} n_{\rm A}, \color{A2color} n_{\rm A_2}, \color{OA2color} n_{\rm OA_2}, \color{Ocolor} n_{\rm O}, \color{OB2color} n_{\rm OB_2}, \color{B2color} n_{\rm B_2}, \color{Bcolor} n_{\rm B} \color{Ocolor})\), we pick out of the \(p\) tensor the probability of that particular microstate.
Pictorially,
\begin{equation}
  p = \raisebox{-0.5\height}{\includegraphics[width=200pt]{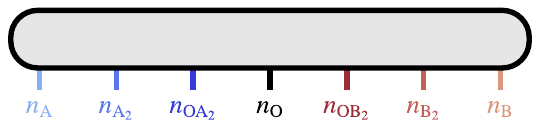}},
  \label{eq:punfactorized}
\end{equation}
gives the probability of state \(\mathbf{n}\) if the index for each dangling leg is the copy number for the corresponding species.
The high-order tensor \(p\) suffers from the curse of dimensionality, but we can build a more tractable order seven tensor from the contraction of smaller-order tensors, one per species:

\begin{align}
  \nonumber & \raisebox{-0.5\height}{\includegraphics[width=200pt]{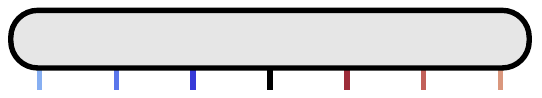}} \\
  & \ \ \ \ \ \ \ \approx \raisebox{-0.5\height}{\includegraphics[width=200pt]{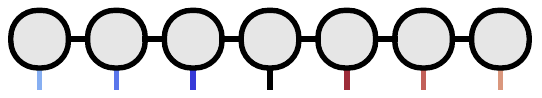}}.
  \label{eq:pfactorized}
\end{align}
This one-dimensional chain of tensors is a factorization into a matrix product state (MPS) or tensor train (TT).
For a sufficiently large bond dimension, the steady-state distributions could always be written in this factorized form, but the bond dimension would need to grow very large.
The computational advantage emerges because real problems appear to involve steady-state distributions which can be very well approximated by an MPS with a modest bond dimension.
In the quantum mechanical context, this ability to compress is well justified in terms of the so-called area laws~\cite{orus2014practical, cramer2006entanglement, srednicki1993entropy, plenio2005entropy, vidal2003entanglement}.
In the CME context, we are unaware of comparatively satisfactory formal justifications for the observed ability to massively truncate the bond dimension, but our GTS calculations in this work required a maximum bond dimension no greater than 100 (see also Appendix~\ref{app:fourth}).
Before discussing those results, we must address how, for a particular choice of bond dimension, we can fit the tensor elements of each MPS tensor to best approximate \(\left| \pi\right>\).
Doing so requires that we first demonstrate how to systematically convert the CME's elementary reactions into a TN form.

\subsection{Rate Operators in a Matrix Product Form}
\label{sec:MPO}

Elements of the rate matrix \(\mathbb{W}\) are specified by both the initial microstate {\bf n} and the final microstate {\bf n'}.
In our TN notation, a rate operator thus takes the form:
\begin{equation}
  \mathbb{\hat{W}} = \raisebox{-0.5\height}{\includegraphics[width=200pt]{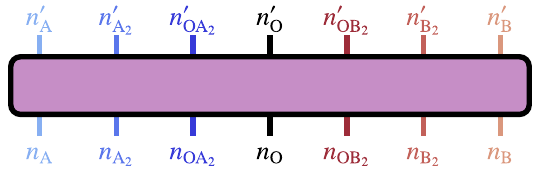}},
  \label{eq:Wunfactorized}
\end{equation}
This \(\mathbb{\hat{W}}\) is very high-order, so it is appealing to try to factorize it into lower-order tensors, as we did with the approximation for the steady-state distribution:
\begin{align}
  \nonumber & \raisebox{-0.5\height}{\includegraphics[width=200pt]{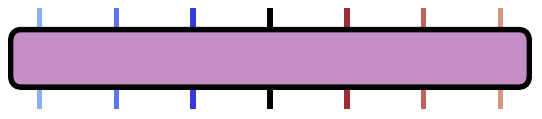}} \\
  & \ \ \ \ \ \ \ \approx \raisebox{-0.5\height}{\includegraphics[width=200pt]{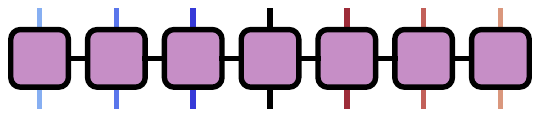}}.
  \label{eq:Wfactorized}
\end{align}
The factorized form is a matrix product operator (MPO).
Each of the rounded rectangles acts on the physical copy number space (colored legs) of a single species, tracking the rates for increasing or decreasing the copy number of that species.
But the species do not change in isolation.
Take, for example, our first reaction, \(2 \color{Acolor} \text{A} \color{Ocolor} \xrightharpoonup{c_1^+} \color{A2color}\text{A}_2\).
This reaction will simultaneously decrease \(\color{Acolor} n_{\rm A}\) by two while increasing \(\color{A2color} n_{\rm A_2}\) by one.
This correlation is naturally expressed in a second-quantized representation in terms of a raising operator \(x_l^\dagger\) that increases the copy number of species \(l\) by one and a lowering operator \(x_l\) that lowers it by one.
For example, the ``gain'' term of the rate operator is simply \(c_1^+ \color{Acolor} x_{\rm A}^2 \color{A2color} x_{\rm A_2}^\dagger\), which acts on the two species in conjunction.
The Doi-Peliti formalism systematically converts elementary reactions into a contribution to the second-quantized rate operator, including both the ``gain'' terms and the ``loss'' terms~\cite{doi1976stochastic,peliti1985path}.
The second-quantized version of the GTS model is discussed further in Appendix \ref{app:first} of this work and Appendix C of Ref.~\cite{strand2024from}.

The black horizontal bond between the first two tensors in the factorization of Eq.~\eqref{eq:Wfactorized} allows us to build an operator that correlates the action on the two different species.
In the case of reaction 1, the correlations are built into a single horizontal bond between neighboring tensors, but the bond dimension only needs to grow linearly to pass correlations down the chain.
Consequently, the TN representation of the rate operator \(\mathbb{\hat{W}}\) can be very efficient from the perspective of both the memory needed to store the operator and the computational effort to perform operations on it.
When we first introduced the rate matrix \(\mathbb{W}\), we conceived of it as a very large matrix with dimension equal to the number of microstates.
Even for a sparse matrix, storing that object in memory grows problematic.
With the factorization, we see that the rate operator \(\mathbb{\hat{W}}\) could be alternatively expressed by storing the values of seven factorized tensors that have order 3 or 4.
Building those tensors from the second-quantized operator is essentially a solved problem \cite{hubig2017generic} that is implemented in tensor network software packages.
Our calculations used the autoMPO function of the ITensor software package to turn Eq.~\eqref{eq:dp} into the MPO~\cite{10.21468/SciPostPhysCodeb.4, 10.21468/SciPostPhysCodeb.4-r0.3}.

\section{Steady-state distributions via DMRG}
\label{sec:DMRG}

Section~\ref{sec:TNs} showed how to construct the seven tensors of the MPO from elementary reactions and their associated rates.
The section furthermore introduced the idea that an MPS factorization could approximate a distribution over microstates.
It remains to compute the values of those MPS tensors which approximate a particular distribution, the steady-state.
The key is to recognize that for the irreducible CME, the steady state is the unique largest eigenvector of \(\hat{\mathbb{W}}\), which must have eigenvalue zero.
In terms of the TNs, we therefore need to solve for the values of the circular tensors to yield:
\begin{equation}
  \raisebox{-0.4\height}{\includegraphics[width=200pt]{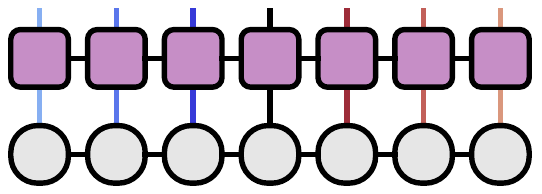}} = \mathbf{0}.
  \label{eq:eigenvalue}
\end{equation}
where $\mathbf{0}$ represents an order-7 tensor of the appropriate dimensions in which each entry is zero.
This equation can be viewed as an optimization problem, where the optimized circular tensors will approximate \(\pi\).

It is computationally taxing to perform operations on order-7 tensors like Eq.~\eqref{eq:punfactorized}, but Eq.~\eqref{eq:pfactorized}'s factorization offers a smarter route.
We want to perform operations on single-species circular tensors, updating them sequentially to solve the global optimization problem of Eq.~\eqref{eq:eigenvalue}.
In the quantum mechanical context, the Density Matrix Renormalization Group (DMRG) algorithms are popular and effective methods to find the extremal eigenvector (ground state) of quantum many-body systems by sweeping through the chain performing optimizations on factorized tensors~\cite{white1992density,white1993density,schollwock2011density}.
It is tempting to simply apply the same algorithms to the CME, but we must remember that the CME is typically non-Hermitian whereas usual quantum mechanical problems seek the ground state of a Hermitian Hamiltonian.

\subsection{Ground states of non-Hermitian operators}
There has been some activity trying to utilize DMRG for non-Hermitian systems.
One approach is to construct a Gram matrix \(\hat{G} = \mathbb{\hat{W}}^\dagger \mathbb{\hat{W}}\), where \(^\dagger\) here denotes a conjugate transpose.
\(\hat{G}\) and \(\mathbb{\hat{W}}\) share the same right zero eigenvector, but \(\hat{G}\) is Hermitian even if \(\mathbb{\hat{W}}\) is non-Hermitian~\cite{cui2015variational, casagrande2021analysis}.
Applying the usual DMRG algorithms to \(\hat{G}\) therefore offers a valid route to construct \(\pi\) for the non-Hermitian GTS, but there is a drawback.
The convergence rate of DMRG is sensitive to the spectral gap between the first and second eigenvalues.
The closer the gap is to zero, the more problematic DMRG convergence becomes. The eigenvalues of $\hat{G}$ are not those of $\mathbb{\hat{W}}$ but instead are the singular values squared of $\mathbb{\hat{W}}$. Then collections of small singular value gaps for $\mathbb{\hat{W}}$ can lead to even smaller gaps in $\hat{G}$~\cite{cui2015variational}.
Another na\"{i}ve method of restoring Hermiticity would be to use the operator $\mathbb{\hat{W}} + \mathbb{\hat{W}^\dagger}$ instead of $\mathbb{\hat{W}}$.
However, this operator will no longer have the same ground state as $\mathbb{\hat{W}}$, thus requiring a more complicated cost function to restore the existence of a variational principle~\cite{guo2022variational}.
A third route is to develop alternative DMRG algorithms that are specifically built for non-Hermitian operators \cite{chan2005density, huang2011biorthonormal, verstraete2004matrix, zhong2024density}.
This route, which holds long-term promise, cannot immediately utilize all of the software tools already developed for Hermitian DMRG.

There are, however, reports of success simply applying the usual DMRG procedure to the non-Hermitian operator~\cite{zhang2020skin,liao2023density,shen2024non}.
We have similarly found that we can compute \(\pi\) for our GTS model by applying a na\"{i}ve DMRG scheme to our non-Hermitian \(\mathbb{\hat{W}}\).
One might anticipate that DMRG would remain stable if there is a small parameter introducing a non-Hermitian perturbation on top of a Hermitian operator.
In our case, one cannot argue that \(\mathbb{\hat{W}}\) is particularly close to Hermitian, so we sought a justification to understand why the convergence works in practice.

\subsection{Rayleigh-Ritz justification for DMRG}
The usual justification for the DMRG algorithm originates with the Rayleigh-Ritz variational principle.
An MPS with a modest bond dimension cannot exactly represent the ground state in general, but the MPS could nevertheless be expanded as a superposition of the exact right eigenvectors \(\ket{\phi_i}\): \(\ket{\Phi_{\rm MPS}} = \sum_i \alpha_i \ket{\phi_i}\).
The conjugate transpose of that MPS can also be expanded as a superposition of the exact left eigenvectors \(\bra{\phi_i}\): \(\bra{\Phi_{\rm MPS}} = \sum_i \beta_i \bra{\phi_i}\).
When \(\mathbb{\hat{W}}\) is non-Hermitian, \(\alpha_i\) and \(\beta_i\) are not simply related to each other, but when \(\mathbb{\hat{W}}\) is Hermitian, they are complex conjugates, \(\alpha_i = \beta_i^*\).
For a Hermitian rate operator \(\mathbb{\hat{W}}\), the Rayleigh-Ritz quotient is a weighted average of \(\mathbb{\hat{W}}\)'s eigenvalues, \(\lambda_i\), so the quotient must fall between the extremal eigenvalues \(\lambda_{\rm min}\) and \(\lambda_{\rm max} = 0\):
\begin{equation}
  \lambda_{\rm min} \leq \frac{\sum_i \abs{\alpha_i}^2 \lambda_i}{\sum_{i} \abs{\alpha_i}^2} \leq 0.
  \label{eq:RRinequality}
\end{equation}
For the Rayleigh-Ritz quotient to approach 0, \(\ket{\Phi_{\rm MPS}}\) must approach a state with \(\alpha_i\) equal to 1 for the ground state and 0 for all other eigenstates.
That \(\ket{\Phi_{\rm MPS}}\) is parameterized by many tensor elements defining the tensors of each single-species tensor of the MPS.
The Rayleigh-Ritz principle tells us that a perturbation to each of those single-species tensors which succeeds in moving the Rayleigh-Ritz quotient closer to zero will have also moved \(\ket{\Phi_{\rm MPS}}\) closer to the true ground state.

There are many flavors of DMRG algorithm that iteratively pass through those MPS tensors to perform perturbations that nudge the Rayleigh-Ritz quotient upward toward zero.
For example, one can attempt to optimize a single-species tensor (one gray circle) while holding the other six single-species tensors fixed.
The six fixed tensors can be contracted with the MPO to produce a new effective operator that acts on the gray circle to be optimized.
Optimization of the single-species tensor is performed efficiently as an Arnoldi~\cite{arnoldi1951principle} maximum eigenvector calculation of that low-rank effective operator.
The DMRG algorithm then moves to the next gray circle down the chain, optimizing it subject to all the others being fixed.
These sweeps through the MPS tend to increase the Rayleigh-Ritz quotient, which therefore pushes \(\ket{\Phi_{\rm MPS}}\) closer to the ground state.
Many more advanced DMRG algorithms exist, some of which incorporate noise \cite{white2005density}, subspace expansion \cite{hubig2015strictly}, or optimizations of more than one tensor at a time \cite{white1992density, white1993density, schollwock2011density}.
For example, our results utilize a two-site DMRG algorithm that sweeps through the MPS doing local maximal eigenvector calculations on the joint space given by two neighboring gray circles.

\subsection{Time evolution justification for DMRG}
Making \(\mathbb{\hat{W}}\) non-Hermitian does not fundamentally disrupt the idea that the global optimization of Eq.~\eqref{eq:eigenvalue} can be decomposed into a sequence of local optimizations, but it does fundamentally alter the Rayleigh-Ritz inequality of Eq.~\eqref{eq:RRinequality}.
Without the relationship between \(\alpha_i\) and \(\beta_i\), the Rayleigh-Ritz quotient is
\begin{equation}
  \frac{\sum_i \alpha_i \beta_i \lambda_i}{\sum_{i} \alpha_i \beta_i},
\end{equation}
and \(\alpha_i \beta_i\) may no longer be a positive real number.
As a consequence, the Rayleigh quotient might now rise above \(\lambda_{\rm max} = 0\), meaning perturbations to \(\ket{\Phi_{\rm MPS}}\) which yield a Rayleigh-Ritz quotient closer to zero might not correspond to states which are closer to the ground state.
One can perform the same DMRG sweeps\textemdash construct an effective operator for a one- or two-site tensor, perform a maximum eigenvector calculation, sweep to the next sites, and iterate\textemdash but without the assurances of the Rayleigh-Ritz principle it is unclear if such an algorithm actually targets the ground state for a non-Hermitian operator.

Here, we provide a different argument that supports that na\"{i}ve application of DMRG to our particular non-Hermitian problems.
Specifically, we note that the Perron-Frobenius theorem implies that any probability distribution over microstates tends to the unique \(\ket{\pi}\) in the long-time limit, that is \(\left|\pi\right> = \lim_{t \to \infty} e^{\mathbb{\hat{W}} t} \ket{p(0)}\).
In essence, this long-time propagation is the solution of the maximal eigenvector by the power method.
The propagation for infinite time can be thought of as an infinite number of discrete timesteps of length \(\Delta t\),
\begin{equation}
  \ket{\pi} = e^{\mathbb{\hat{W}} \Delta t} \hdots e^{\mathbb{\hat{W}} \Delta t} \ket{p(0)}.
\end{equation}
For sufficiently small \(\Delta t\), the tensor network implementation of the time-dependent variational principle (TN TDVP) gives a controllable approximation for the time evolution from one MPS into another.
If the initial \(\ket{p(0)}\) is itself an MPS, that TN TDVP algorithm constructs another MPS for the time-propagated state \(\ket{p(\Delta t)}\).
In lieu of the DMRG procedure, one could therefore find \(\ket{\pi}\) by the power method through the sequential application of the TN TDVP algorithm with a sufficiently small \(\Delta t\)~\cite{nicholson2023quantifying}.
This TN TDVP solution for the ground state may require many timesteps, but even without the Rayleigh-Ritz method, Perron-Frobenius guarantees that \(\ket{\pi}\) will eventually be reached.

To make TN TDVP reach the steady-state distribution faster, one could try simply increasing \(\Delta t\), but in a loose sense, this will cause some MPS tensors to ``get out ahead of the others''.
TN TDVP requires matrix exponentials for time \(\Delta t\) using an effective rate operator that acts on one or two sites.
One can think of those steps of the algorithm as propagating the distribution for some sites to time \(t + \Delta t\) while leaving the others behind at time \(t\).
Once the full sweep has been completed, all tensors have been advanced to \(t + \Delta t\).
Within a sweep, some tensors capture behavior at time \(t\) while others have already advanced to \(t + \Delta t\).
For finite \(\Delta t\), a timestep error emerges because an individual tensor is propagated with an effective rate operator built from a mixture of time \(t\) and time \(t + \Delta t\) tensors.
Accurate evolution of the distribution requires the small \(\Delta t\) limit.

Reaching the steady-state distribution, however, does not require accurate evolution.
One could update \(\ket{p}\) in a manner that has significant timestep errors but nevertheless approaches the same fixed point.
We note that for our master equation problems, the fixed point of a TN TDVP sweep does not depend on the timestep \(\Delta t\).
To see this fact, imagine that we first evolved an initial distribution \(\ket{p(0)}\) using TN TDVP with a very small \(\Delta t\) to approximate \(\ket{\pi}\) by the power method.
We could take the final MPS and apply one more step of TN TDVP with a very large \(\Delta t\) timestep.
The old concern about timestep error ceases to matter when all of the site tensors have already reached their steady-state (\(t \to \infty\)) values.
Now, each site tensor will already be a zero eigenvector for its effective rate operator, meaning the time propagation by \(\Delta t\) will not change the site tensor.
In other words, once an MPS representation of \(\ket{\pi}\) is found, further time evolution \emph{with any choice of \(\Delta t\)} will stick at \(\ket{\pi}\).

This observation justifies running TN TDVP steps with large timesteps.
The path to a converged MPS may be less smooth than running TN TDVP with a small timestep, but a large timestep means the MPS can be significantly changed in a single sweep.
Some site tensors could get way out ahead of the others, leading to a thrashing as the site tensors are each propagated based on outdated values of the other tensors.
That thrashing could shrink a radius of convergence, but if we can cleverly seed each calculation to fall within some convergence region, we can be confident that the final stationary state of the TN TDVP algorithm is \(\ket{\pi}\).

Procedurally, the \(\Delta t \to \infty\) limit of a TDVP timestep is particularly nice.
A TDVP timestep ordinarily involves matrix exponentials of an effective rate operator acting for time \(\Delta t\).
In the long-time limit, that matrix exponential is simply related to the maximal eigenvalue and the associated eigenvector.
Indeed, Haegeman et al.\ have noted that the DMRG algorithm can be understood as a special case of the TDVP algorithm in the infinite-timestep limit~\cite{haegeman2016unifying}.
The same DMRG algorithm that targets a Hermitian ground state by a Rayleigh-Ritz justification therefore targets the stationary state \(\ket{\pi}\) of our non-Hermitian rate operator \(\hat{W}\)~\footnote{Here we take the defining step of a DMRG algorithm to be computing a maximal eigenvector of a local operator, not necessarily that such an eigenvector is found using the Davidson algorithm. When we say the DMRG algorithm is unchanged from the Hermitian case, the local problem's maximal eigenvectors must of course be found with an appropriate solver that does not assume a symmetric local operator.}.
Convergence from an arbitrary starting distribution $\ket{p(0)}$ is of course not certain given the thrashing, but it is reasonable to hope that there exists a radius of convergence of a non-negligible size.
The numerical experiments which constitute the main results of this paper confirm this optimistic notion.

\section{Assessing bistability}
\label{sec:bistability}

We know that the GTS model can exhibit bistability for certain values of the elementary rates in the CRN.
There are two different ways to extract information about that bistability from the DMRG calculations.
The first route, which directly utilizes the steady-state distribution $\ket{\pi}$, is a static characterization of the bimodality of the stationary distribution.
The second route, which requires excited-state DMRG calculations, more directly probes the timescale of dynamics.

\begin{figure}[h]
\centering
\includegraphics[width=0.48\textwidth]{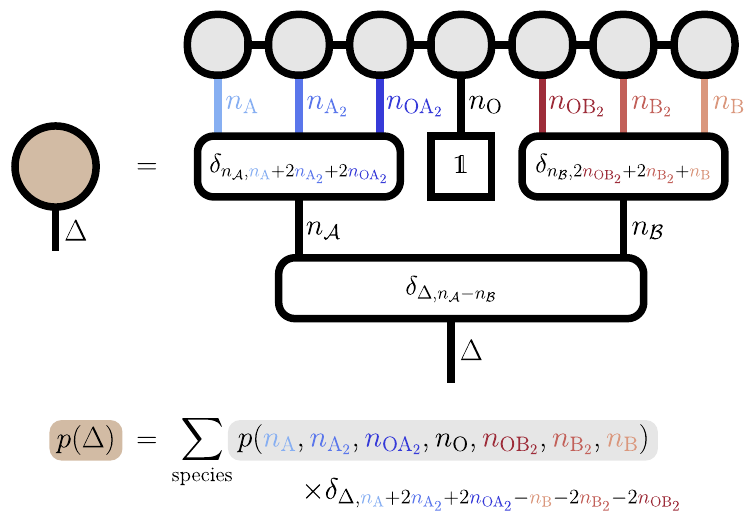}
\caption{\label{fig:Delta}
  DMRG returns a tensor network which assigns a probability to each microstate (gray MPS).
  Summing over those microstates to get the marginal distribution of the order parameter \(\Delta\) requires summing over all microstates that yield the particular value of \(\Delta\).
  As \(n_{\text{O}}\) does not influence \(\Delta\), that coordinate is traced over (contracted with the identity vector $\mathds{1}$).
  Sums over the occupation numbers of the other species are efficiently realized by building a tensor network of Kronecker delta tensors.
  Upon contraction, one is left with an order-one tensor (brown circle) whose index ranges over the allowed values of \(\Delta\) and whose elements are the marginal steady-state probabilities.}
\end{figure}

Let us start by directly using $\ket{\pi}$, found as the DMRG ground state.
The order parameter \(\Delta\) sorts microstates based on how many excess A proteins there are compared to B, so the marginal distribution \(P_{\rm ss}(\Delta)\) shows whether \(\ket{\pi}\) is dominated by A-rich states, B-rich states, or some mixture.
Building \(P_{\rm ss}(\Delta)\) from $\ket{\pi}$ is a relatively straightforward marginalization, one which is particularly easy to compute using tensor networks.
Figure~\ref{fig:Delta} shows how sparse tensors can be constructed from Kronecker deltas to contract over the various species involving protein A, transforming to a new tensor that is indexed by the total number of A, $n_{\mathcal{A}}$.
An equivalent Kronecker delta tensor characterizes the microstates by the total number of B, $n_{\mathcal{B}}$.
A final Kronecker delta tensor, $\delta_{\Delta, n_{\mathcal{A}}-n_{\mathcal{B}}}$, combines all mixtures of $n_{\mathcal{A}}$ and $n_{\mathcal{B}}$ that produce identical values of $\Delta$.
The contraction scheme of Fig.~\ref{fig:Delta} builds $P(\Delta)$ from any $\ket{p}$, but when applied to $\ket{\pi}$ it gives the stationary $P_{\rm ss}(\Delta)$.
The bimodality of that $P_{\rm ss}(\Delta)$ can be assessed from the skewness $\gamma$ and kurtosis $\kappa$ in terms of Sarle's bimodality coefficient~\cite{pfister2013good}
\begin{equation}
{\rm BC} = \frac{\gamma^2 + 1}{\kappa}.
\end{equation}
By construction, this BC ranges from 0 to 1.
Values that exceed the BC of a uniform distribution (5/9) typically indicate a bimodal or multimodal distribution~\cite{pfister2013good}.
In Sec.~\ref{sec:exploration}, we numerically show how that BC varies as a function of elementary GTS rates, building the kinetic phase diagram in Fig.~\ref{fig:main_results}.

The particular shape of $P_{\rm ss}(\Delta)$ cannot simply yield the timescale for switching between A-rich and B-rich states, but bimodality is a good proxy for bistability.
A more direct measure of the bistability requires dynamical timescales, which come from the excited-state eigenvalues of $\hat{\mathbb{W}}$.
Let us order the eigenvalues of $\hat{\mathbb{W}}$ with $0 = \lambda_0  > \lambda_1 > \lambda_2 \hdots$ and with $\ket{\pi} \equiv \ket{\phi_0}$, $\ket{\phi_1}$, $\ket{\phi_2}$, $\hdots$ being the associated right eigenvectors.
An initial probability distribution can be expanded in the eigenbasis $\ket{p(0)} = \sum_i \alpha_i \ket{\phi_i}$.
Time evolution is simple in the eigenbasis:
\begin{equation}
  \ket{p(t)} = e^{\hat{\mathbb{W}}t} \ket{p(0)} = \sum_i \alpha_i e^{\lambda_i t} \ket{\phi_i}
\end{equation}
Because all excited-state $\lambda_i$ values are negative and ordered, the long-time distribution's deviations from the ground state are dominated by the first excited state, $\ket{\phi_1}$:
\begin{equation}
  \lim_{t \to \infty} \ket{p(t)} \approx \ket{\pi} + \frac{\alpha_1}{\alpha_0} \ket{\phi_1} e^{\lambda_1 t},
\end{equation}
meaning $\tau = 1 / \lambda_1$ is the slowest timescale for relaxation.
For two-state kinetics with rate $k_{\mathcal{AB}}$ for switching from state $\mathcal{A}$ to $\mathcal{B}$ and rate $k_{\mathcal{BA}}$ for switching back, the relaxation timescale is $\tau = (k_{\mathcal{AB}} + k_{\mathcal{BA}})^{-1}$.
When the two-state kinetics is symmetric, $k_{\mathcal{AB}} = k_{\mathcal{BA}} = \lambda_1 / 2$.
Hence, the typical time to switch from an A-rich to B-rich state, $\tau_{\mathcal{AB}}$, follows from the eigenvalue of the first excited state:
\begin{equation}
  \label{eq:dmrgtau}
  \tau_{\mathcal{AB}} = \frac{1}{k_{\mathcal{AB}}} = \frac{2}{\lambda_1}.
\end{equation}

By ``projecting out'' the ground-state, one can modify the rate operator $\hat{\mathbb{W}}$ to solve for $\lambda_1$ and $\ket{\phi_1}$ as a ground state of the modified rate operator
\begin{equation}
  \mathbb{\hat{W}}' = \mathbb{\hat{W}} - \xi \ket{\phi_0}\bra{\phi_0},
  \label{eq:modifiedrateoperator}
\end{equation}
where $\xi$ is a constant that shifts the eigenvalue of the original ground state down from 0~\cite{stoudenmire2012studying}.
The penalty $\xi$ is chosen to be sufficiently large that the first excited-state of $\hat{\mathbb{W}}$ becomes the ground state of $\hat{\mathbb{W}}'$.
Because our $\hat{\mathbb{W}}$ is typically not Hermitian, $\ket{\phi_0}$ and $\bra{\phi_0}$ are not typically conjugate transposes of each other.
As we have seen, $\ket{\phi_0} = \ket{\pi}$, and conservation of probability requires that $\bra{\phi_0} = \bra{\mathds{1}}$ is a vector of ones.

\section{Numerical Exploration of the GTS Model}
\label{sec:exploration}

The GTS model is parameterized by eight distinct kinetic parameters, and the bistability is strongly influenced by those parameters.
For any choice of the elementary rates, Gillespie sampling can be run with or without the maximum copy number $M$ truncation.
For example, the model acts as a bistable switch when $c_1^+ = c_1^- = c_2^+ = c_2^- = 5$, $c_\mathrm{A}^+ = c_\mathrm{B}^+ = 1$, $c_\mathrm{A}^- = c_\mathrm{B}^- = 0.25$, in line with prior studies of the model \cite{warren2004enhancement, warren2005chemical}.
Throughout the paper, we will keep the first six rate parameters fixed, focusing only on the dependence on the decomposition rates.
Increasing the decomposition rates to $c_\mathrm{A}^- = c_\mathrm{B}^- = 0.8$ causes the distribution $P_{\rm ss}(\Delta)$ to approach a unimodal one with a rapid relaxation rate $\lambda_1$.
Gillespie sampling offers a useful confirmation that truncation at $M = 20$ does not significantly alter bimodality or bistability, but Gillespie samples must be regenerated for every set of rate parameters that one wishes to study.

\begin{figure}[ht]
\centering
\includegraphics[width=0.5\textwidth]{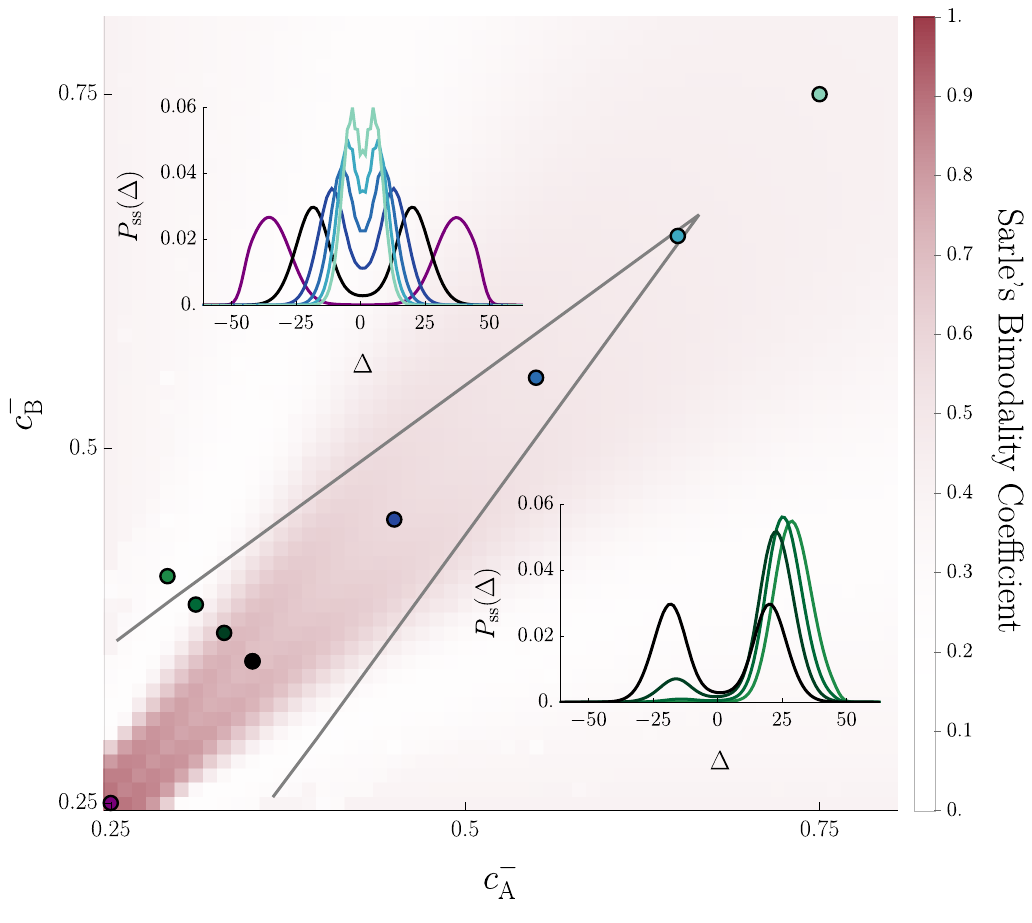}
\caption{\label{fig:main_results}
  A kinetic phase diagram shows the GTS model bimodality as a function of protein decomposition rates.
  Prior mean-field analysis identified a bistable region inside the gray wedge.
  DMRG calculations, repeated over a grid of $c_{\rm A}^-$ and $c_{\rm B}^-$ values yield the steady-state distribution $\ket{\pi}$ and the marginal $P_{\rm ss}(\Delta)$.
  Strongly bimodal $P_{\rm ss}(\Delta)$ distributions arise for kinetic parameters within the gray wedge, but stochastic fluctuations cause the bimodal region to shrink relative to the mean-field prediction.
  Insets highlight two distinct ways the distribution can lose bimodality.
  The top left inset tracks $P_{\rm ss}(\Delta)$ moving along a diagonal ($c_{\rm A}^- = c_{\rm B}^-$), showing a merging of the two peaks at high decomposition rates, akin to a second order phase transition. 
  Note also that the relevance of the discreteness of the copies of proteins is most apparent in the case of high decomposition rates (i.e., the sea green curve).
  The bottom right inset tracks $P_{\rm ss}(\Delta)$ moving perpendicular to that diagonal, showing a first-order-like switch between a unimodal A-dominated distribution and a unimodal B-dominated distribution.
  The kinetic phase diagram is generated with fixed parameters $c_1^+ = c_1^- = c_2^+ = c_2^- = 5$ and $c_\mathrm{A}^+ = c_\mathrm{B}^+ = 1$ using DMRG convergence thresholds described in Appendix~\ref{app:fourth}.}
\end{figure}

Because parameter exploration would require so many repeated Gillespie simulations, prior work has explored whether mean-field analyses can capture the stochastic kinetics~\cite{blossey2008mean}.
For the GTS model, Warren and ten Wolde solved a system of seven differential equations at a negligible computational cost to generate a kinetic phase diagram mapping whether the GTS will act as a bistable switch as a function of the depletion rates of the two proteins, $c_{\mathrm{A}}^-$ and $c_\mathrm{B}^-$.
The analysis, briefly reproduced in Appendix~\ref{app:second}, divides the parameter space (gray line in Fig.~\ref{fig:main_results}) into regions with and without anticipated bistability.
That mean-field analysis is computationally cheap, requiring only minutes on a laptop, but there are limitations.
The mean-field calculations return a single average value for the copy number of each species, so it is not possible to obtain a distribution of outcomes like $P_{\rm ss}(\Delta)$.
Moreover, the mean value will typically be noninteger, an especially egregious limitation for the $\mathrm{O}$, $\mathrm{OA_2}$, and $\mathrm{OB_2}$ species, which in reality always have either one or zero copies present.

\subsection{DMRG Exploration}
Having laid out the TN tools, we are in a position to illustrate how useful they are for the parameter exploration that has previously been studied with sampling and mean-field methods.
By repeating DMRG calculations of the ground state on a grid of $c_{\rm A}^-$ and $c_{\rm B}^-$ values, we computed Sarle's bimodality coefficient for each ground state distribution.
Figure~\ref{fig:main_results} shows that strongly bimodal steady states roughly align with the mean-field calculation, though there are some notable deviations from the mean-field analysis.
The bimodality coefficient demonstrates that incorporating fluctuations narrows the bimodal regime for low decomposition rates, as compared with the mean-field analysis.
Graphically, the misidentification of the bimodal areas is identified by the white patches located inside the gray wedge of Fig.~\ref{fig:main_results}.
This deviation from mean field is further visualized in the lower-right inset of Fig.~\ref{fig:main_results}, where we plot how $P_{\rm ss}(\Delta)$ responds to a simultaneous decrease of one decomposition rate and increase of the other.
Moving from black to green, $P_{\rm ss}(\Delta)$ shows the A-rich states growing at the expense of the B-rich, with the two humps equal in height for the black $c_{\rm A}^- = c_{\rm B}^- = 0.35$ distribution.
We can additionally view how $P_{\rm ss}(\Delta)$ changes when the decomposition rates are increased in concert.
The upper-left inset of Fig.~\ref{fig:main_results} shows the two distinct humps at positive and negative \(\Delta\) merging into each other at \(\Delta = 0\), much like a second-order phase transition in an Ising model.

\begin{figure}
\centering
\includegraphics[width=0.5\textwidth]{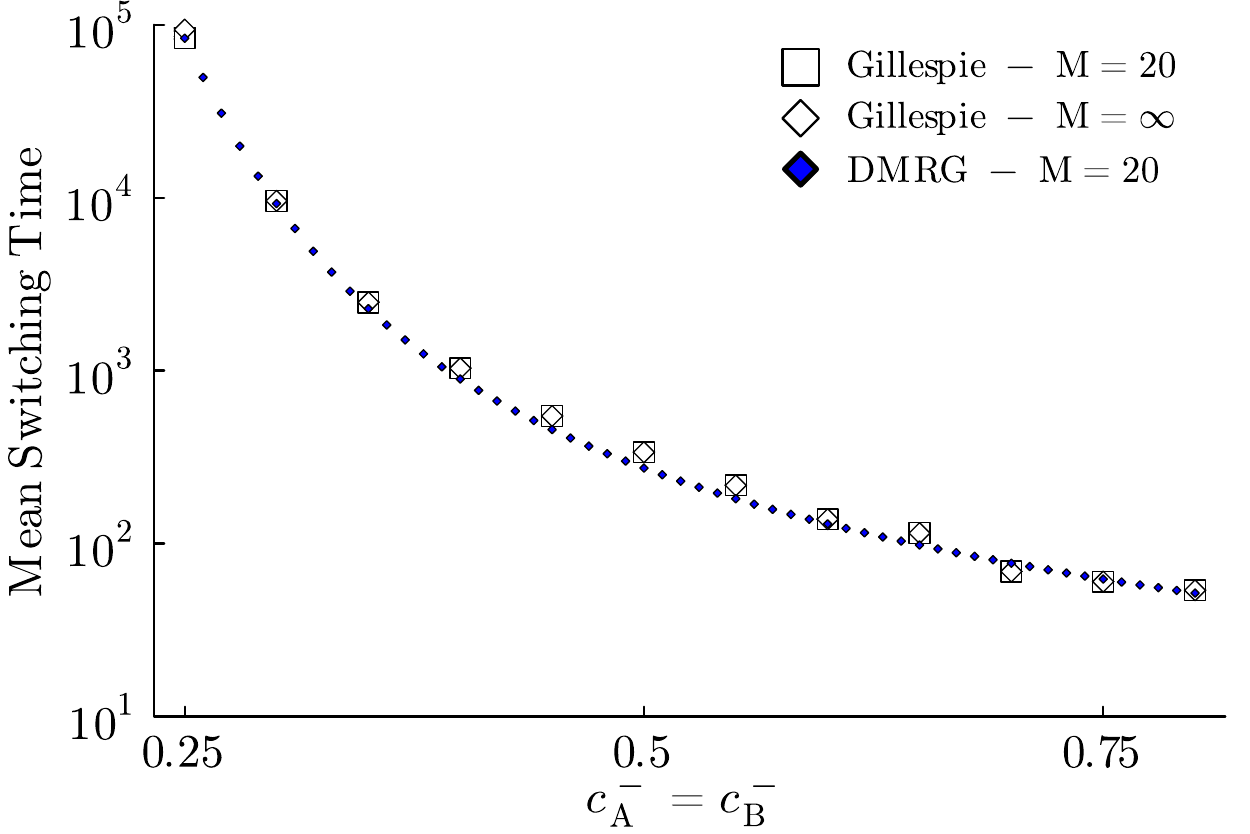}
\caption{\label{fig:switching_rates}
  Along the diagonal of Fig.~\ref{fig:main_results}, $c_{\rm A}^- = c_{\rm B}^-$ and $P_{\rm ss}(\Delta)$ supports two peaks.
  The timescale for switches between positive and negative $\Delta$ spans orders of magnitude.
  That timescale is estimated from long Gillespie trajectories with or without a maximal occupation number (\(M = 20\) and \(M = \infty\)), showing that the finite truncation at $M = 20$ only barely begins to impact the switching time for the smallest decomposition rates of $c_\mathrm{A}^- = c_\mathrm{B}^- = 0.25$.
  Excited-state DMRG calculations reproduce the $M = 20$ timescale via Eq.~\ref{eq:dmrgtau}.
  Gillespie estimates of switching time require defining two basins with a threshold $d$, one with $\Delta \geq d$ and the other with $\Delta \leq -d$.
  The mean switching time is the average time between switches, averaged over $10^9$ units of time.
  The mean switching time is weakly impacted by the chosen threshold $d$, and an appropriately chosen threshold must shift as a function of decomposition rate (see the top left inset of Fig.~\ref{fig:main_results}).
  Those shifting basin definitions create the impression that the Gillespie estimates are slightly noisy about the DMRG curve.} 
\end{figure}

Even for parameters lying outside the mean-field-theory bistable region, e.g., $c_{\rm A}^- = c_{\rm B}^- = 0.75$, $P_{\rm ss}(\Delta)$ retains a bimodal character.
That light blue $P_{\rm ss}(\Delta)$ reveals two humps merging so close to \(\Delta = 0\) that stochastic fluctuations readily enable flips between A-rich and B-rich states.
This set of parameters requires us to be cautious about differentiating between bimodality of the steady-state distribution and bistability, with the latter carrying dynamical information about switching from one set of states at one time to another set at a later time.
That bistability is better probed, not from the bimodality coefficient, but rather from the slowest timescale of relaxation, \(\lambda_1\), coming from the excited-state DMRG calculations.
Figure~\ref{fig:switching_rates} plots $\tau_{\mathcal{AB}}$ as a function of the decomposition rate, explicitly showing that when $P_{\rm ss}(\Delta)$ has a small ``barrier'' separating the two peaks, interconversion between positive and negative \(\Delta\) is rapid.
When performing excited-state DMRG with $c_{\rm A}^- = c_{\rm B}^-$, as in Fig.~\ref{fig:switching_rates}, we find some numerical advantage when using a penalty method with a modified rate operator that differs slightly from the $\hat{\mathbb{W}}'$ of Eq.~\eqref{eq:modifiedrateoperator} (see Appendix~\ref{app:third}).

\subsection{Sampling versus DMRG}

Although we generated Fig.~\ref{fig:main_results} using the DMRG calculations, the figure could also be generated with sufficient Gillespie sampling.
For every set of rate parameters, new Gillespie trajectories could be generated to sample $P_{\rm ss}(\Delta)$ and to estimate $\tau_{\mathcal{AB}}$.
We close the paper by reporting on performance in a manner that should help illustrate conditions for which DMRG is worth it and conditions for which sampling clearly makes more sense.
The conversation is necessarily nuanced.
One important reason why the discussion cannot be as simple as a comparison of timing of the two calculations is that the comparison is between apples and pears.
While DMRG calculations were used to extract $P_{\rm ss}(\Delta)$, the DMRG was actually converged to yield the more fine-grained $\ket{\pi}$.
To estimate the steady-state probability for \emph{every} microstate from Gillespie sampling is clearly impractical, so the DMRG is always computing an object with extra information.
To the extent that the extra information is relevant to the question at hand, DMRG is always ``worth it''.

If, however, one is only running DMRG to get $\ket{\pi}$ as a means to $P_{\rm ss}(\Delta)$, one must ask whether that $P_{\rm ss}(\Delta)$ should have just been sampled.
Even this question has an annoying amount of nuance.
The answer depends on whether the rate parameters are in a slow- or fast-switching regime.
It depends on whether one intends to sample with advanced sampling tricks or in an approximate way with, for example, $\tau$-leaping.
And it especially depends on whether one aims to learn $P_{\rm ss}(\Delta)$ for a single set of rate parameters or for a collection of many parameters as we did in Fig.~\ref{fig:main_results}.
We focus on the last case, assuming that the goal is that parameter scan.

\begin{figure}
\centering
\includegraphics[width=0.5\textwidth]{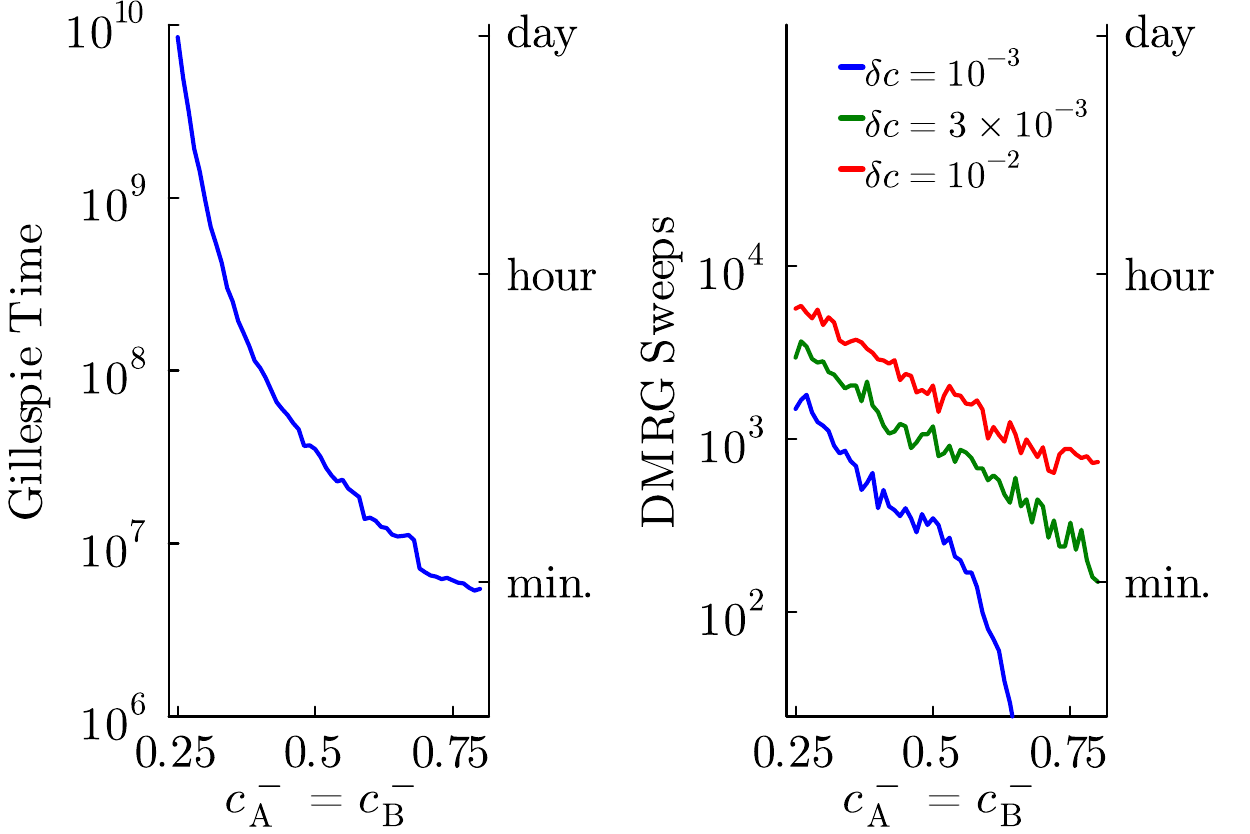}
\caption{\label{fig:timing_figure} 
  Converging $P_{\rm ss}(\Delta)$ from Gillespie samples requires many switching events, and the Gillespie time needed to generate 100,000 switching events grows dramatically longer as the decomposition rates $c_{\rm A}^- = c_{\rm B}^-$ decrease.
  An advantage of DRMG is that a converged DMRG calculation can seed the calculation for new decomposition rates.
  Starting with a converged $\ket{\pi}$ at one value of decomposition rates, we report the number of DMRG sweeps needed to converge and find $\ket{\pi}$ for a new value of decomposition rates which have been decreased by a step size of $\delta c$.
  By starting at large $c_{\rm A}^- = c_{\rm B}^-$ and decreasing with a small $\delta c$, one can more cheaply construct $P_{\rm ss}(\Delta)$ down through the rare-event regime.
  Wall time is essentially proportional to the length of the Gillespie time and the number of DMRG sweeps.
  To guide the eye, an approximate conversion for the single CPU calculations is provided on the right axis, using the fact that a single DMRG sweep took about 0.4 seconds and 100,000 units of Gillespie time could be simulated in about 1 second of wall time. 
}   
\end{figure}

Sampling offers the advantage that each set of rate parameters is an independent calculation, so these calculations are trivially parallelized.
For each value of the rate parameters, we estimate $\tau_{\mathcal{AB}}$ by averaging over Gillespie trajectories with 1,000 switching events.
The cost to converge $P_{\rm ss}(\Delta)$ depends on the degree of convergence one hopes to achieve.
On the way to convergence, the relative probabilities of the positive- and negative-$\Delta$ peaks will not be exactly equal.
A coin flip calculation reflects that each peak will be $50 \pm 3\%$ after $10^3$ switching events, $50 \pm 1\%$ after $10^4$ switching events, and $50 \pm 0.3\%$ after $10^5$ switching events.
We define that final threshold as our converged Gillespie sampling and estimate the time to converge by $10^5 \times \tau_{\mathcal{AB}}$, plotted in Fig.~\ref{fig:timing_figure}.
As anticipated, the rare-event regime with small $c_{\rm A}^-$ and $c_{\rm B}^-$ is dramatically more expensive to exhaustively sample by brute force Gillespie.

DMRG can offer computational advantages, particularly in this rare-events regime.
Quantitatively assessing the benefit is challenging because any benefit depends on the particular way we choose to leverage one DMRG calculation to seed the next.
The essential strategy is to start by converging DMRG in a corner of the parameter space.
It has been noted that the convergence time of DMRG typically scales inversely with the spectral gap of the system \cite{verstraete2006matrix}, making it advantageous to begin with the fast-switching parameters $c_\mathrm{A}^- = c_\mathrm{B}^- = 0.8$.
Having converged $\ket{\pi}$ for those rates, we should perturb the rates by a small step and reconverge.
The expense of such a scheme is dominated by how many DMRG sweeps are necessary for each reconvergence.
If you take small steps in parameter space, you have to reconverge many times.
If you take large steps in parameter space, however, each reconvergence could grow much more expensive because the new $\ket{\pi}$ could be significantly different from the seed.
To quantify these effects, we report in Fig.~\ref{fig:timing_figure} the number of DMRG sweeps necessary to reconverge $\ket{\pi}$ at $(c_{\rm A}^- - \delta c, c_{\rm B}^- - \delta c)$ when seeded with the converged MPS for $\ket{\pi}$ at $(c_{\rm A}^-, c_{\rm B}^-)$ for three different step sizes: $\delta c = 10^{-2}, 3 \times 10^{-3}, $ and $10^{-3}$.
To evaluate when the DMRG sweeps should terminate, we first ran an expensive, conservative convergence with extra DMRG sweeps.
We then calculated the 1-norm between that ``converged'' distribution and the converging distribution, terminating the DRMG sweeps once the 1-norm dropped below $3 \times 10^{-3}$.
Technical details of the convergence are discussed in Appendix~\ref{app:fourth}, but here we highlight that our 1-norm threshold is chosen so that the $P_{\rm ss}(\Delta)$ from DMRG is about as converged as a histogram from trajectories of length $10^5 \times \tau_{\rm AB}$.

Figure~\ref{fig:timing_figure} shows that the expense of both types of calculations depends on the parameter regime, with fast decomposition rates always being easy.
DMRG, however, scales better as the problem starts to become hard.
While the cost to converge brute-force Gillespie histograms varies over three orders of magnitude, the number of DMRG sweeps varies over only one.
That DMRG expense is really the marginal cost of getting the next distribution by dragging the last one over to the new rate parameters.
How hard it is to perform that reconvergence depends on how dramatically $\ket{\pi}$ changes as a function of the rate parameters.
Taking smaller steps $\delta c$, naturally results in smaller changes in $\ket{\pi}$ and fewer required sweeps.

\section{Discussion}
\label{sec:discussion}

Ours is not the first work to suggest that the raising and lowering operator techniques (Doi-Peliti) are a powerful way to analyze the chemical master equation~\cite{grassberger1980fock, cardy1985epidemic, van1998wilson, sasai2003stochastic, lan2006variational, walczak2009stochastic, weber2017master, vastola2021solving}, nor is it the first to apply tensor network tools to CRNs~\cite{ hegland2010numerical, kazeev2014direct, dolgov2015simultaneous, liao2015tensor, gelss2016solving, vo2017adaptive, helms2019dynamical, ion2021tensor, prugger2023dynamical, dolgov2024tensor, merbis2024effective}.
Our aim here has been to unite these two ideas, using the GTS model to demonstrate the systematic workflow from elementary reactions to Doi-Peliti representations to matrix product operators to steady-states.
We believe that the case for approximating $\ket{\pi}$ without sampling is strongest for sensitivity analysis.
Reseeding DMRG calculations with $\ket{\pi}$ for nearby parameters provides a natural way to reuse information from previously solved distributions, affording a computational advantage when one hopes to build a kinetic phase diagram like Fig.~\ref{fig:main_results}.
Sensitivity analysis in the form of gradients should be even more easily computed if one measures the small changes to $\ket{\pi}$ in response to a small $\delta c$.
Recently, a differentiable Gillespie algorithm has been developed to compute those gradients from sampling~\cite{rijal2024differentiable}.
It will be interesting to compare and contrast in the future.

The GTS model has a modest enough state space that its steady state can actually be found with sparse-matrix solvers \cite{strand2024from}.
We used that fact to aid in the convergence analysis (see Appendix~\ref{app:fourth}), but we hope the TN tools will extend far beyond seven chemical species.
TN tools are routinely used to study quantum spin chains with hundreds of spins, so it is not unreasonable to think that DMRG will be able to numerically investigate large CRNs with tens or even hundreds of species.
Contrasting the quantum spin chains with the CRNs, however, highlights some important challenges that remain.

Firstly, large quantum spin chains often involve sites with low physical dimension (spins are up or down), whereas our chemical species have occupation numbers ranging from 0 to $M$.
That difference grows the size of the MPS tensors, meaning 100 chemical species is a much harder problem than 100 spins.
Clever ways to reduce the physical dimension will be important. 
One possibility would be to use quantized tensor trains (QTTs), which break down each physical dimension into the digits of its binary form, thus potentially reducing the scaling of storage cost in terms of the sizes of the individual state spaces from linear to logarithmic~\cite{kazeev2014direct}. 
Another possibility could be to find a better set of basis functions to represent the distributions~\cite{kryven2015solution, prugger2023dynamical, cai2024modified}.
This approach would be analogous to the way the problem of large physical dimensionality is confronted in quantum chemistry \cite{chan2005density}. 
Notably, it has already been shown that the Doi-Peliti formalism can be expressed in terms of Charlier or Hermite polynomials \cite{ohkubo2012one}. 
Finally, for systems not exhibiting ``discreteness-induced multimodality,'' it may prove advantageous to use the CFPE instead of the CME, as done by Liao et al.~\cite{liao2015tensor}, which would allow one to use a discretization grid of one's choosing.

Secondly, unlike in spin chains, there is usually no notion of chemical species being connected along a one-dimensional chain.
In applying the MPS ansatz to the GTS model, we are not assuming that species only interact with their neighbor in the chain, but correlations between distant species must be passed down the chain.
Capturing those correlations requires a larger bond dimension than if the strongly interacting species had been connected to each other for the TN ansatz.
As long as calculations with a large enough bond dimension can be performed, it is not necessary to optimize the topology of the TN ansatz, but it stands to reason that it could be advantageous to approximate the joint steady-state distribution by a TN whose connectivity mirrors the connectivity of the chemical species in the CRN. 
For technical reasons, tensor networks with loops lack an easily accessible ``canonical form''~\cite{Tindall2023Gauging}, frustrating the ability to fit a loopy tensor network via DMRG.
By breaking apart loops in the CRN's natural graph, one could instead approximate the joint distributions with tree tensor networks, which are amenable to DMRG~\cite{nakatani2013efficient}.
Recent progress developing techniques for the ``regauging'' of loopy tensor networks~\cite{evenbly2018gauge, alkabetz2021tensor, tindall2024efficient, haghshenas2019conversion, wang2024tensor, evenbly2024loop} offers hope that a DMRG-style approach may eventually be possible with a TN that matches the CRN topology.

Finally, while we successfully applied the ordinary DMRG algorithm for the GTS model, its performance was more delicate than we would have liked.
Unless we seeded the distributions well, we experienced numerical instability.
Appendix~\ref{app:fourth} describes a slightly complicated reseeding scheme that we needed to engineer to confirm that DMRG sweeps did not get lost far from a converged MPS.
We hope that DMRG algorithms specifically tailored for non-Hermitian systems~\cite{verstraete2004matrix, zhong2024density} will help improve the robustness and speed of convergence.

\section{Acknowledgments}
We thank Cathryn Murphy and Nils E.\ Strand for helpful discussions.
The material presented in this manuscript is based upon work supported by the National Science Foundation under Grant No.\ 2239867.
T.R.G.\ was additionally supported by the Camille Dreyfus Teacher-Scholar Program.

\appendix

\section{The Doi-Peliti Formalism}
\label{app:first}

The Doi-Peliti formalism turns a CME into its second-quantized representation, meaning it expresses elementary reactions in terms of the raising and lowering operators:
\begin{align}
x_l^\dagger \ket{n_l} &= \ket{n_l + 1} \\
x_l \ket{n_l} &= n_l\ket{n_l - 1} \\
x_l \ket{0} &= 0.
\end{align}
The subscripts indicate on which sites the operator acts, with $l$ being $\mathrm{A}$, $\mathrm{A_2}$, $\mathrm{OA_2}$, $\mathrm{O}$, $\mathrm{OB_2}$, $\mathrm{B_2}$, or $\mathrm{B}$ in the case of the GTS model.
When chemical species \(l\) is capped at a maximum occupation number of $M$, the operators take the form of $M+1$ by $M+1$ matrices:
\begin{equation}
\begin{split}
    x_l=\begin{pmatrix}
    0 & 1 & 0 & \cdots & 0 & 0\\
    0 & 0 & 2 & \cdots & 0 & 0\\
    0 & 0 & 0 & \cdots & 0 & 0\\
    \vdots & \vdots & \vdots & \ddots & \vdots & \vdots\\
    0 & 0 & 0 & \cdots & 0 & M\\
    0 & 0 & 0 & \cdots & 0 & 0\\
    \end{pmatrix}
\end{split}
\begin{split}
    \text{ and }
\end{split}
\begin{split}
    x_l^\dagger=\begin{pmatrix}
    0 & 0 & 0 & \cdots & 0 & 0\\
    1 & 0 & 0 & \cdots & 0 & 0\\
    0 & 1 & 0 & \cdots & 0 & 0\\
    \vdots & \vdots & \vdots & \ddots & \vdots & \vdots\\
    0 & 0 & 0 & \cdots & 0 & 0\\
    0 & 0 & 0 & \cdots & 1 & 0\\
    \end{pmatrix}.
\end{split}
\end{equation}
In Appendix C of Ref.~\cite{strand2024from}, we have shown that it is convenient to additionally define matrices
\begin{equation}
\begin{split}
    y_l=\begin{pmatrix}
    1 & 0 & \cdots & 0 & 0 & 0\\
    0 & 1 & \cdots & 0 & 0 & 0\\
    \vdots & \vdots & \ddots & \vdots & \vdots & \vdots\\
    0 & 0 & \cdots & 1 & 0 & 0\\
    0 & 0 & \cdots & 0 & 1 & 0\\
    0 & 0 & \cdots & 0 & 0 & 0\\
    \end{pmatrix}
\end{split}
\begin{split}
    \text{ and }
\end{split}
\begin{split}
    z_l=\begin{pmatrix}
    1 & 0 & \cdots & 0 & 0 & 0\\
    0 & 1 & \cdots & 0 & 0 & 0\\
    \vdots & \vdots & \ddots & \vdots & \vdots & \vdots\\
    0 & 0 & \cdots & 1 & 0 & 0\\
    0 & 0 & \cdots & 0 & 0 & 0\\
    0 & 0 & \cdots & 0 & 0 & 0\\
    \end{pmatrix},
\end{split}
\end{equation}
which are the identity provided one is not within one (or two) counts of the maximum occupancy $M$.
Reaction 1 in Fig.~\ref{fig:overall_scheme} involves the loss of two A coupled to the production of a dimer A$_2$.
One can demonstrate that reaction 1 brings to $\mathbb{\hat{W}}$ a contribution
\begin{equation}
  \hat{\mathbb{W}} = c_1^+\Big(x_{\text{A}}^2 x_{\text{A}_2}^\dagger - x_{\text{A}}^{\dagger 2} x_{\text{A}}^2 y_{\text{A}_2}\Big),
\end{equation}
where the positive part is the gain term and the negative part the loss term~\cite{strand2024from}.
As is typical, we adopt a shorthand whereby only non-identity terms acting on $\ket{\mathbf{n}}$ are explicitly specified.
The identities are implicit.
For example,
\begin{equation}
  x_{\mathrm{A}}^2 x_{\mathrm{A_2}}^\dagger =x_{\mathrm{A}}^2 \otimes x_{\mathrm{A_2}}^\dagger \otimes \mathbb{I}_{\mathrm{OA_2}}\otimes \mathbb{I}_{\mathrm{O}}\otimes \mathbb{I}_{\mathrm{OB_2}}\otimes \mathbb{I}_{\mathrm{B_2}}\otimes \mathbb{I}_\mathrm{B}
\end{equation}

While operators acting on A, A$_2$, B, and B$_2$ are $M+1$ by $M+1$ matrices, those acting on OA$_2$, O, and OB$_2$ are only 2 by 2 matrices because our \(n_{\rm DNA} = 1\) manifold restricts the maximum occupancy of those species to 1.
Note that for species involving O, our prior definitions of the operators become:
\begin{equation}
  x_l=\begin{pmatrix}
  0 & 1 \\
  0 & 0
  \end{pmatrix}, \ 
  x_l^\dagger=\begin{pmatrix}
  0 & 0 \\
  1 & 0
  \end{pmatrix},
  \ \text{and} \ 
  y_l=\begin{pmatrix}
  1 & 0 \\
  0 & 0
  \end{pmatrix}.
\end{equation}
Separately carrying out the Doi-Peliti procedure on all fourteen elementary reactions yields:

\begin{widetext}
\begin{equation}
\begin{minipage}{0.5\textwidth}
\begin{align*}
    & \hat{\mathbb{W}}_1 = c_1^+\Big(x_{\text{A}}^2 x_{\text{A}_2}^\dagger - x_{\text{A}}^{\dagger 2} x_{\text{A}}^2 y_{\text{A}_2}\Big) \\
    & \hat{\mathbb{W}}_2 = c_1^-\Big(x_{\text{A}}^{\dagger 2} x_{\text{A}_2} - z_{\text{A}} x_{\text{A}_2}^\dagger x_{\text{A}_2}\Big) \\
    & \hat{\mathbb{W}}_3 = c_2^+\Big(x_{\text{A}_2} x^\dagger_{\text{OA}_2}x_{O} - x_{\text{A}_2}^\dagger x_{\text{A}_2} y_{\text{OA}_2} x_{\text{O}}^\dagger x_{\text{O}} \Big) \\
    & \hat{\mathbb{W}}_4 = c_2^- \Big( x_{\text{A}_2}^\dagger x_{\text{OA}_2} x_{\text{O}}^\dagger - x_{\text{OA}_2}^\dagger x_{\text{OA}_2} y_{\text{O}} y_{\text{A}_2} \Big) \\
    & \hat{\mathbb{W}}_5 = c_\mathrm{A}^+\Big(x_{\text{A}}^\dagger x_{\text{O}}^\dagger x_{\text{O}} - y_{\text{A}} x_{\text{O}}^\dagger x_{\text{O}} \Big) \\
    & \hat{\mathbb{W}}_6 = c_\mathrm{A}^+\Big(x_{\text{A}}^\dagger x_{\text{OA}_2}^\dagger x_{\text{OA}_2} - y_{\text{A}} x_{\text{OA}_2}^\dagger x_{\text{OA}_2} \Big) \\
    & \hat{\mathbb{W}}_7 = c_\mathrm{A}^-\Big(x_\text{A}-x_{\text{A}}^\dagger x_{\text{A}}\Big)
\end{align*}
\end{minipage}
\hfill
\begin{minipage}{0.5\textwidth}
\begin{align}
    & \nonumber \hat{\mathbb{W}}_8 = c_1^+\Big(x_{\text{B}}^2 x_{\text{B}_2}^\dagger - x_{\text{B}}^{\dagger 2} x_{\text{B}}^2 y_{\text{B}_2}\Big) \\
    & \nonumber \hat{\mathbb{W}}_9 = c_1^-\Big(x_{\text{B}}^{\dagger 2} x_{\text{B}_2} - z_{\text{B}} x_{\text{B}_2}^\dagger x_{\text{B}_2}\Big) \\
    & \nonumber \hat{\mathbb{W}}_{10} = c_2^+\Big(x_{\text{B}_2} x^\dagger_{\text{OB}_2}x_{O} - x_{\text{B}_2}^\dagger x_{\text{B}_2} y_{\text{OB}_2} x_{\text{O}}^\dagger x_{\text{O}} \Big) \\
    & \nonumber \hat{\mathbb{W}}_{11} = c_2^- \Big( x_{\text{B}_2}^\dagger x_{\text{OB}_2} x_{\text{O}}^\dagger - x_{\text{OB}_2}^\dagger x_{\text{OB}_2} y_{\text{O}} y_{\text{B}_2} \Big) \\
    & \nonumber \hat{\mathbb{W}}_{12} = c_\mathrm{B}^+\Big(x_{\text{B}}^\dagger x_{\text{O}}^\dagger x_{\text{O}} - y_{\text{B}} x_{\text{O}}^\dagger x_{\text{O}} \Big) \\
    & \nonumber \hat{\mathbb{W}}_{13} = c_\mathrm{B}^+\Big(x_{\text{B}}^\dagger x_{\text{OB}_2}^\dagger x_{\text{OB}_2} - y_{\text{B}} x_{\text{OB}_2}^\dagger x_{\text{OB}_2} \Big) \\
    & \nonumber \hat{\mathbb{W}}_{14} = c_\mathrm{B}^-\Big(x_\text{B}-x_{\text{B}}^\dagger x_{\text{B}}\Big).
\end{align}
\end{minipage}
\label{eq:dp}
\end{equation}
\end{widetext}

The rate operator $\hat{\mathbb{W}}$ for the full CRN is simply the sum of the individual contributions from each reaction:

\begin{equation}
\hat{\mathbb{W}} = \sum_{i=1}^{14} \hat{\mathbb{W}}_i.
\end{equation}

\section{The Mean Field Approach}
\label{app:second}

As a point of comparison, we replicated the mean-field results presented in \cite{warren2004enhancement}.
In the large-system-size limit, the mass-action kinetics form of the 14 elementary reactions gives seven coupled differential equations for the mean concentration of the species.
Steady-state kinetics requires that the concentration of each species has become stationary, so

\begin{align}
0 &= \frac{d[\mathrm{A}]}{dt} = -2c_1^+[\mathrm{A}]^2 + 2c_1^-[\mathrm{A_2}] + c_\mathrm{A}^+([\mathrm{O}] + [\mathrm{OA_2}]) - c_\mathrm{A}^-[\mathrm{A}] \notag \\
0 &= \frac{d[\mathrm{A_2}]}{dt} = c_1^+[\mathrm{A}]^2 - c_1^-[\mathrm{A_2}] - c_2^+[\mathrm{O}][\mathrm{A_2}] + c_2^-[\mathrm{OA_2}] \notag \\
0 &= \frac{d[\mathrm{OA_2}]}{dt} = c_2^+[\mathrm{O}][\mathrm{A_2}] - c_2^-[\mathrm{OA_2}] \notag \\
0 &= \frac{d[\mathrm{O}]}{dt} = -c_2^+[\mathrm{O}]([\mathrm{A_2}] + [\mathrm{B_2}]) + c_2^-([\mathrm{OA_2}] + [\mathrm{OB_2}]) \notag \\
0 &= \frac{d[\mathrm{OB_2}]}{dt} = c_2^+[\mathrm{O}][\mathrm{B_2}] - c_2^-[\mathrm{OB_2}] \notag \\
0 &= \frac{d[\mathrm{B_2}]}{dt} = c_1^+[\mathrm{B}]^2 - c_1^-[\mathrm{B_2}] - c_2^+[\mathrm{O}][\mathrm{B_2}] + c_2^-[\mathrm{OB_2}] \notag \\
0 &= \frac{d[\mathrm{B}]}{dt} = -2c_1^+[\mathrm{B}]^2 + 2c_1^-[\mathrm{B_2}] + c_\mathrm{B}^+([\mathrm{O}] + [\mathrm{OB_2}]) - c_\mathrm{B}^-[\mathrm{B}].
\label{eq:mean_field}
\end{align}

We selected two different initial conditions, \((\color{Acolor} n_{\rm A}, \color{A2color} n_{\rm A_2}, \color{OA2color} n_{\rm OA_2}, \color{Ocolor} n_{\rm O}, \color{OB2color} n_{\rm OB_2}, \color{B2color} n_{\rm B_2}, \color{Bcolor} n_{\rm B} \color{Ocolor}) = (10, 10, 1, 0, 0, 0, 0)\) and \((0, 0, 0, 0, 1, 10, 10)\), the first A-rich and the second B-rich.
From these seeds, we numerically integrated the mass action kinetics with the Euler scheme using $\delta t = 0.01$ until the time derivatives of all species concentrations had fallen below $10^{-10}$.
If one trajectory resulted in $\Delta > 1$ while the other resulted in $\Delta < -1$, the elementary rate parameters were considered to be in a bistable regime.
We scanned a grid of $c_{\rm A}^-$ and $c_{\rm B}^-$ values, identifying the bistable points then plotted in Fig.~\ref{fig:main_results} the convex hull of those bistable parameters, a plot that aligns with Fig.\ 2 of~\cite{warren2004enhancement}.

\section{Excited-state DMRG for Symmetric Decomposition Rates}
\label{app:third}
We find that, for symmetric parameter sets ($c_{\rm A}^- = c_{\rm B}^-$), the penalty-method calculation of the first excited state is even more robust when DMRG is applied to a modified rate operator with a slightly different penalty (one that would have been natural for a Hermitian problem):
\begin{equation}
\mathbb{\hat{W}}'' = \mathbb{\hat{W}} - \xi \ket{\pi}\bra{\pi},
\label{eq:excited_state_DMRG}
\end{equation}
where $\bra{\pi}$ is now the conjugate transpose of $\ket{\pi}$.
In general, the maximal right eigenvector of this $\mathbb{\hat{W}}''$ is not the targeted first excited state of $\mathbb{\hat{W}}$, $\ket{\phi_1}$.
When $c_{\rm A}^- = c_{\rm B}^-$, we show (next paragraph) that $\bra{\pi} \ket{\phi_1} = 0$.
Eq.~\eqref{eq:excited_state_DMRG} then implies $\hat{\mathbb{W}}''\ket{\phi_1} = \hat{\mathbb{W}}\ket{\phi_1}$, so $\ket{\phi_1}$ is an eigenstate of the original rate operator and the new modified operator.
Because $\bra{\pi}\ket{\pi}$ does not vanish, that $\ket{\phi_1}$ becomes the ground state of $\hat{\mathbb{W}}''$ for a sufficiently large $\xi$.

Completing the argument requires us to prove that $\bra{\pi}\ket{\phi_1} =0$ when $c_{\rm A}^- = c_{\rm B}^-$.
We do so with a symmetry consideration, namely $\bra{\pi}$ is symmetric under the exchange of the protein labels $\mathrm{A}$ and $\mathrm{B}$ while $\ket{\phi_1}$ is antisymmetric under the same exchange.
The inner product between symmetric and antisymmetric vectors necessarily vanishes.
To be slightly more explicit, if \((\color{Acolor} n_{\rm A}, \color{A2color} n_{\rm A_2}, \color{OA2color} n_{\rm OA_2}, \color{Ocolor} n_{\rm O}, \color{OB2color} n_{\rm OB_2}, \color{B2color} n_{\rm B_2}, \color{Bcolor} n_{\rm B} \color{Ocolor}) = (n_1, n_2, n_3, n_4, n_5, n_6, n_7)\) has steady-state probability $p$, so too must microstate \((n_7, n_6, n_5, n_4, n_3, n_2, n_1)\).
In contrast, $\ket{\phi_1}$'s antisymmetry is required because the orthonormality of $\mathbb{\hat{W}}$'s left and right eigenvectors gives $\bra{\phi_0}\ket{\phi_1} = \bra{\mathds{1}}\ket{\phi_1} = 0$.
For the sum of the elements in $\ket{\phi_1}$ to be zero, microstates \((n_1, n_2, n_3, n_4, n_5, n_6, n_7)\) and \((n_7, n_6, n_5, n_4, n_3, n_2, n_1)\) have equal magnitude and opposite sign.

\section{DMRG convergence}
\label{app:fourth}

\subsection{Ground state scan}
To scan the kinetic parameter space and construct Fig.~\ref{fig:main_results}, DMRG calculations had to be repeatedly initiated with seeds, and then DMRG sweeps were run until a convergence threshold was reached.
The sweeps were performed using two-site DMRG with a maximum bond dimension of 100 and a singular value cutoff of $10^{-15}$.
Throughout the full scan of the parameter space, the singular value cutoff was the limiting factor\textemdash the largest bond dimension of a converged ground state MPS was 54.
Because we have non-Hermitian rate operators, in place of a symmetric solver like the Davidson algorithm, the Arnoldi method~\cite{arnoldi1951principle} was used with a maximum Krylov space dimension of 3 to solve the local eigenvalue problems. 

We sought to initiate the parameter scan in the easiest-to-compute corner of Fig.~\ref{fig:main_results}.
Since DMRG convergence tends to be faster when spectral gaps are larger, we thus started with the largest values of $c_{\rm A}^-$ and $c_{\rm B}^-$.
There, the DMRG readily converges when seeded from a uniform distribution.
That convergence is monitored most simply by tracking the ``energy'', $\bra{\phi_\mathrm{trial}}\hat{\mathbb{W}}\ket{\phi_\mathrm{trial}}/\braket{\phi_\mathrm{trial}}{\phi_\mathrm{trial}}$, where $\ket{\phi_{\rm trial}}$ is the MPS state and $\bra{\phi_{\rm trial}}$ is the complex conjugate.
We refer to this quantity as the energy because it would be the expected energy if $\hat{\mathbb{W}}$ were a Hermitian Hamiltonian.
Even for the non-Hermitian rate operator, the quantity makes sense to inspect because it would vanish for the true ground state $\ket{\pi}$, at which $\hat{\mathbb{W}} \ket{\pi} = 0$.
We considered $10^{-5}$ to be an appropriate threshold for identifying the ground state.
For $c_{\rm A}^- = c_{\rm B}^- = 0.8$, this level of convergence was reached from the uniform seed within $10^4$ DMRG sweeps.

After converging the steady-state distribution in the top right corner of Fig.~\ref{fig:main_results}, we wanted to scan in such a manner that the steady-state distribution would move as gently as possible.
The gentlest move came from simultaneously decreasing both decomposition rates by the same small step, so we reseeded the DMRG convergences by passing the converged distribution at $c_{\rm A}^-$ and $c_{\rm B}^-$ down the diagonal to $c_{\rm A}^- - \delta c$ and $c_{\rm B}^- - \delta c$.
At the boundaries of the scanned region, horizontal and vertical steps were taken, as necessary, to fully scan a grid of values $c_{\rm A}^-$ and $c_{\rm B}^-$ from 0.25 to 0.8 in increments of 0.01.
DMRG can be reconverged with steps in the elementary rates of size 0.01 (see Fig.~\ref{fig:timing_figure}), but we preferred to take those steps via 200 smaller steps ($\delta c = 5 \times 10^{-5}$).
In practice, we would run 100 DMRG sweeps for each new parameter and took the MPS with the lowest energy.
Notice that we did not need to solve for the ground state on a $5 \times 10^{-5}$ grid to benefit from the smaller step sizes.
We only needed to connect points on the 0.01 grid by finely discretized lines along which the ground state distribution was ``adiabatically dragged''.

We found that the DMRG would occasionally get knocked off course, finding a distribution so far from the ground state that a very large number of sweeps was required to converge.
To improve the stability of the scan, we used $P_{\rm ss}(\Delta)$ as a cheap-to-compute indicator of severe numerical instabilities.
For example, if the MPS for a point on the grid with spacing 0.01 had an energy exceeding $10^{-4}$ or gave a marginal distribution $P_{\rm ss}(\Delta)$ with negative or complex values, it was clear that the MPS had veered too far from $\ket{\pi}$.
In these cases, we would attempt up to $10^4$ DMRG sweeps to reconverge to below the energy threshold of $10^{-5}$.
Occasionally, even those extra sweeps were insufficient. 
If the sweeps had been sufficient to at least return a MPS corresponding to a $P_{\rm ss}(\Delta)$ without negative or complex values, the MPS with the lowest absolute energy during the extra sweeps was used. 
If even this was not achieved, we returned the MPS with the last $\delta c = 5 \times 10^{-5}$ grid point with positive real $P_{\rm ss}(\Delta)$ probabilities.

\begin{figure}
\centering
\includegraphics[width=0.5\textwidth]{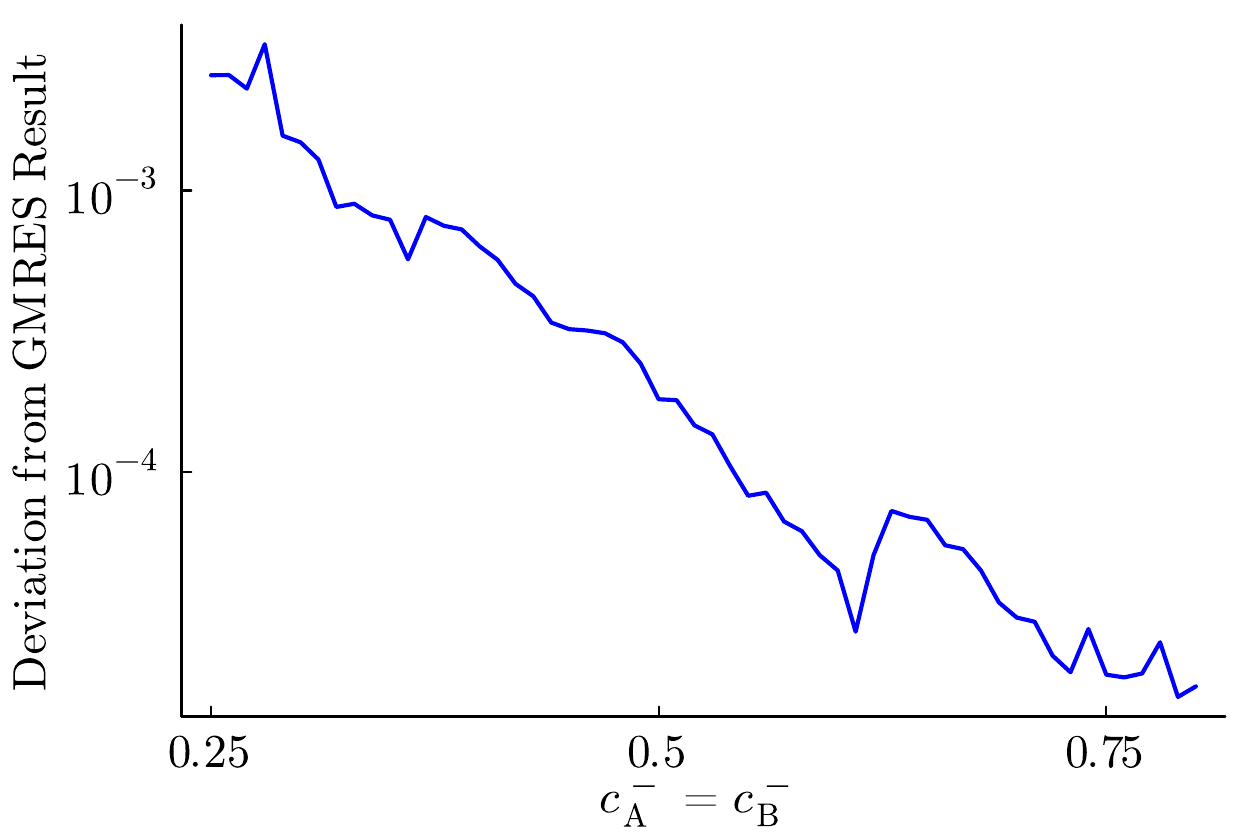}
\caption{\label{fig:GMRES_figure}
 The deviations in 1-norm of the joint steady-state probability distributions obtained from DMRG (with singular value cutoffs of $10^{-15}$) from the results obtained from GMRES. The distributions are normalized to have a probabilistic interpretation (i.e., 1-norms are 1.0).}
\end{figure}

As the energy gets closer to zero, it is a reflection that a TDVP evolution of the MPS state would move by smaller and smaller amounts.
Arrested dynamics, as reflected by a tiny ``energy'', is necessary at the steady state, but dynamics could become very slow even far away from the ground state.
A more thorough validation of the DMRG convergence is offered by a comparison with a sparse linear algebra solver that does not use the compression of the MPS ansatz.
Because the GTS state space is small enough~\cite{strand2024from}, we used GMRES to converge $\ket{\pi^{\rm GMRES}}$ for $c_{\rm A}^- = c_{\rm B}^-$.
Fig.~\ref{fig:GMRES_figure} plots the 1-norm of the difference $\ket{\pi^{\rm GMRES}} - \ket{\pi^{\rm DMRG}}$, reflecting that our ``energy'' threshold indeed returns a very good approximation to the ground state.
Summing over all microstates, the absolute value of the greatest deviation between the GMRES probability and the DMRG probability we observed was around $3 \times 10^{-3}$.
Suppose one summed only over the errors in the positive-$\Delta$ microstates.
As a worst-case scenario, the $c_{\rm A}^- = c_{\rm B}^-$ DMRG ground state would over- or under-state the positive-$\Delta$ probability as $0.5 \pm 3 \times 10^{-3}$, a level of convergence that requires the $10^5$ Gillespie switching events we discussed in the main text.

Comparisons with GMRES furthermore showcase how the MPS approximation acts as a type of data compression.
The matrix product states obtained are intended to be low-parametric functions capable of returning an accurate estimate of the probability associated with each of the 583,433 possible microstates in the model.
Thus, it is of interest to compare the number of parameters (i.e., tensor elements) in the matrix product states obtained with the number of microstates.
Among the 3,136 possible parameter sets considered, the number of tensor elements required to fit the distribution ranged from 4,718 to 34,893.
This corresponds to between 0.8\% and 6.0\% of the number of microstates in the model.
The median number of required tensor elements was 12,909, or 2.2\% of the number of microstates.
These results indicate that the MPS ansatz was usually able to provide a roughly 50-fold compression of the data representing the joint probability distribution.
We believe that once we scale our approach to larger systems and use tensor networks with topologies more directly mimicking the topology of the system's dynamics, it will be possible to obtain much larger compression ratios.

\subsection{Timing calculations}

For the experiments concerning convergence times, small steps were taken from each of the parameter sets with $c_\mathrm{A}^- = c_\mathrm{B}^-$ on a grid with a spacing of 0.01 between 0.25 and 0.8 to the corresponding parameter set with both $c_\mathrm{A}^-$ and $c_\mathrm{B}^-$ reduced by some step $\delta c$. For each such step, 20,000 sweeps at a cutoff of $10^{-15}$ were provided for complete convergence. Every 10 sweeps a copy of the MPS was saved. It was then possible to track the convergence with respect to the number of sweeps by monitoring the deviation in 1-norm between the distribution obtained at a given number of sweeps and the final distribution obtained after 20,000 sweeps. All distributions were normalized to have a probabilistic interpretation (i.e., a 1-norm of 1.0). The number of sweeps for convergence was defined to be the minimum number of sweeps required to reduce the 1-norm deviation to the fully converged distribution to below $3 \times 10^{-3}$. To obtain the data required to produce Fig.~\ref{fig:timing_figure}, this procedure was then repeated for $\delta c = 10^{-2}, 3 \times 10^{-3}, $ and $10^{-3}$. The number of sweeps required for the $\delta c = 10^{-3}$ series falls off for larger values of $c_\mathrm{A}^- = c_\mathrm{B}^-$ as the induced change in 1-norm due to the step taken eventually falls below $3 \times 10^{-3}$ in that region of parameter space. One can gain confidence in the degree of convergence after 20,000 sweeps by comparing the change in the 1-norm deviation to the final distribution in the final 10 sweeps with the change resulting from the 10 sweeps directly after our threshold of $3 \times 10^{-3}$ had been reached. We found that the changes in 1-norm deviation from the final distribution for the last 10 sweeps were always less than 10 \% (and in fact almost always less than 1 \%) of those resulting from the 10 sweeps directly after ``convergence'' had been reached.

\subsection{Excited state scan}

Our reseeding scheme could also be directly generalized to solve for excited states along the diagonal where $c_\mathrm{A}^-=c_\mathrm{B}^-$. For every small step in parameter space, $\ket{\pi}$ was updated first using $\hat{\mathbb{W}}$ for the appropriate kinetic parameters. Then $\ket{\phi_1}$ was updated using $\hat{\mathbb{W}}''$ as defined in Eq. (\ref{eq:excited_state_DMRG}) using the same kinetic parameters and the $\ket{\pi}$ just found. Both $\ket{\pi}$ and $\ket{\phi_1}$ were reseeded such that the $\ket{\pi}$ at the previous parameter set was used as the seed for the calculation of $\ket{\pi}$ for the current parameter set and the $\ket{\phi_1}$ at the previous parameter set was used as the seed for the calculation of $\ket{\phi_1}$ for the current parameter set. However, to obtain accurate switching times from excited-state DMRG in the rare-events regime, it was necessary to further increase the accuracy used. Thus, the singular cutoff was reduced to $10^{-20}$, and a further 10,000 sweeps were provided at each grid point to ensure full convergence of the excited states at this higher accuracy.

\bibliography{reference.bib}

\end{document}